\documentclass[journal=jpcafh,manuscript=article]{achemso}
\usepackage{graphicx}
\usepackage{amsmath}
\usepackage{amssymb}
\usepackage{multirow}
\usepackage{numprint}
\usepackage{siunitx}
\usepackage{subfig}
\usepackage{adjustbox}
\usepackage{appendix}
\usepackage{mathtools}

\title{Overcoming Artificial Multipoles in Intramolecular Symmetry-Adapted Perturbation Theory}

\author{Du Luu and Konrad Patkowski}
\affiliation{Department of Chemistry and Biochemistry, Auburn University,
Auburn, AL 36849}
\email{patkowsk@auburn.edu}

\date{\today}

\begin{document}

\begin{abstract}
Intramolecular symmetry-adapted perturbation theory (ISAPT) is a method to compute and decompose the noncovalent interaction energy between two molecular fragments \textbf{A} and \textbf{B} covalently connected via a linker \textbf{C}.
However, the existing ISAPT algorithm displays several issues for many fragmentation patterns, including an artificially repulsive electrostatic energy (even when the fragments are hydrogen-bonded) and very large and mutually cancelling induction and exchange-induction terms.
We attribute those issues to the presence of artificial dipole moments at the interfragment boundary, as the atoms of \textbf{A} and \textbf{B} directly connected to \textbf{C} are missing electrons on one of their hybrid orbitals.
Therefore, we propose several new partitioning algorithms which reassign one electron, on a singly occupied link hybrid orbital, from \textbf{C} to each of \textbf{A}/\textbf{B}.
Once the contributions from these link orbitals are added to fragment density matrices, the computation of ISAPT electrostatic, induction, and dispersion energies proceeds exactly a normal, and the exchange energy expressions need only minor modifications.
Among the link partitioning algorithms introduced, the so-called ISAPT(SIAO1) approach (in which the link orbital is obtained by a projection onto the intrinsic atomic orbitals (IAOs) of a given fragment followed by orthogonalization to this fragment's occupied space) leads to reasonable values of all ISAPT corrections for all fragmentation patterns, and exhibits fast and systematic basis set convergence.
This improvement is made possible by a significant reduction in magnitude (even though not a complete elimination) of the unphysical dipole moments at the interfragment boundaries.
We demonstrate the utility of the improved ISAPT partitioning by examining intramolecular interactions in several pentanediol isomers, examples of linear and branched alkanes, and the open and closed conformations of a family of N-arylimide molecular torsion balances.
\end{abstract}

\maketitle 

\section{Introduction}\label{sec:intro}

Weak noncovalent interactions are ubiquitous in nature as they occur both between separate molecules
and between nonbonded fragments of the same molecule. As a result, the knowledge and understanding of 
these interactions is indispensable in many areas of chemistry, physics, materials science, and 
biology. When the energy of a weak interaction is computed using quantum chemistry, it is highly
desirable to use a computational protocol that is both quantitative and conceptual, that is,
it provides an accurate overall value and its meaningful decomposition into individual terms that
can be separately interpreted and understood. There are many approaches that satisfy one of these
conditions but only very few methods can do both; among the latter, symmetry-adapted
perturbation theory (SAPT) \cite{Jeziorski:94,Szalewicz:05a} is a particularly attractive choice 
thanks to its many robust variants and efficient computer implementations. The interaction
energy in SAPT is decomposed into four major contributions: electrostatics (Coulomb interaction
of unperturbed molecular charge densities, including charge penetration), 
exchange (the short-range repulsion stemming from
the Pauli exclusion principle), induction (the
polarization of one molecule by the interacting partner), and dispersion (the correlation between
instantaneous charge density fluctuations).
For example, at the lowest (qualitatively accurate) SAPT0
level of theory, the interaction energy is approximated as a sum of the following corrections, grouped together to account for the four major contributions:
\begin{equation}
E_{\rm int}^{\rm SAPT0}=\left( E^{(10)}_{\rm elst} \right)+
\left( E^{(10)}_{\rm exch} \right) + \left(
E^{(20)}_{\rm ind,resp}+E^{(20)}_{\rm exch-ind,resp}+
\delta_{\rm HF} \right) + \left( 
E^{(20)}_{\rm disp}+E^{(20)}_{\rm exch-disp} \right)
\label{eq:sapt0}
\end{equation}
In Eq.~(\ref{eq:sapt0}) and below, the consecutive superscripts denote orders of perturbation theory with respect to intermolecular interaction and intramolecular correlation (thus, there is no intramolecular correlation in SAPT0).
Furthermore, the additional subscript ``resp'' denotes the relaxation (response) of each molecule's Hartree-Fock (HF) orbitals to the electric field of the interacting partner, and the $\delta_{\rm HF}$ term brings in higher-order induction
and exchange-induction effects contained in the supermolecular HF interaction energy.
SAPT has been employed to compute and interpret intermolecular
interactions for numerous complexes of theoretical and experimental interest; many of these 
applications, together with major advances in the SAPT methodology, have been summarized in the
recent reviews \cite{Szalewicz:12,Hohenstein:12,Jansen:14,Patkowski:20,Garcia:20}.

While there are many options to decompose the intermolecular interaction energy, and several ways to do the same thing for a covalent bond \cite{Mitoraj:09,Levine:16,Levine:17},
the physical decomposition of intramolecular noncovalent interactions is much less understood.
While it is not at present possible to quantify the {\em entire} nonbonded interaction involving all molecular
fragments, if one can identify two nonbonded fragments \textbf{A} and \textbf{B} connected
to each other through a linker fragment \textbf{C}, there are a few options to define and partition
the interaction energy between \textbf{A} and \textbf{B} in the presence of \textbf{C}. On one hand,
one could remove \textbf{C} and cap the dangling bonds of \textbf{A} and \textbf{B} with hydrogen
atoms. As the interaction in question is now intermolecular, one can use any of the standard SAPT
variants to evaluate its contributions \cite{Meitei:16,Meitei:17}. However, the fragmentation
approach requires altering the molecular system and, if the linking fragment is small, artificial
means to avoid an overly pronounced repulsion between the capping hydrogens. Two alternatives
that do not require cutting and capping have been introduced by Gonthier and coworkers 
\cite{Gonthier:14,Parrish:15,Pastorczak:15}.
The first one involves localizing the occupied orbitals on the
fragments \textbf{A}, \textbf{B}, \textbf{C} and constructing
a non-Hermitian zeroth-order Hamiltonian \cite{Gonthier:14} in
which the \textbf{A}--\textbf{B} interactions are removed
following the Chemical Hamiltonian approach introduced by
Mayer \cite{Mayer:83}. These removed interactions are then
brought back as the perturbation that gives rise to the
noncovalent intramolecular interaction \cite{Pastorczak:15}.
The need for a non-Hermitian, biorthogonal formalism makes 
the theory somewhat more complicated, and the resulting energy 
decomposition is a little different than in standard SAPT:
the polarization corrections cannot be separated from their
exchange counterparts (for example, the entire first-order
energy is a single term), but the second-order energy 
includes an explicit delocalization (charge transfer) term
in addition to the induction and dispersion ones.

The second approach \cite{Parrish:15} leads to the same
set of corrections as intermolecular SAPT (at its lowest,
SAPT0 level). In this method, referred to as ISAPT,
the occupied orbitals are 
localized according to the intrinsic bond orbital (IBO)
scheme \cite{Knizia:13}. Subsequently, the
HF wavefunctions for \textbf{A} and \textbf{B} 
are obtained by solving the HF equations
for each fragment embedded in the HF potential of linker 
\textbf{C} (that is, the \textbf{A}--\textbf{C} and
\textbf{B}--\textbf{C} interactions are included but the
\textbf{A}--\textbf{B} one is not). The product of the
\textbf{A} and \textbf{B} wavefunctions obtained in this way 
can then be used as the zeroth-order wavefunction for a 
standard SAPT expansion.

In any variant, the treatment of the interfragment boundaries
is critical to the performance of the method. In the default
ISAPT protocol, the doubly occupied IBOs describing the 
interfragment bonds
are treated as part of the linker \textbf{C} (it is assumed
that (\textbf{A},\textbf{C}) and (\textbf{B},\textbf{C})
are each connected by a single bond only). To maintain 
zero net charges on \textbf{A} and \textbf{B} (as long as the
entire molecule is electrically neutral), one proton worth of 
the charge on the atom from \textbf{A} (\textbf{B}) that
takes part in the interfragment bond is reassigned to 
\textbf{C}. Thus, no spurious charges are created at the
interfragment boundaries (obviously, the presence of such 
charges would badly skew the SAPT electrostatic energy).
However, the same cannot be said about spurious dipole moments,
even when the interfragment bonds are nonpolar. Imagine,
for illustration, that the fragmentation is performed for
the propane molecule to calculate the noncovalent interaction
between two CH$_3$-- groups connected by the --CH$_2$-- 
linker. Then, each of the fragments \textbf{A} and \textbf{B}
features a central carbon atom, with its nuclear charge 
reduced by one, and three hydrogen atoms located along the
directions of three $sp^3$ hybrid orbitals. However, as the
fourth hybrid orbital that would complete the tetrahedral
symmetry is missing, such a fragment has a substantial
dipole moment, and we might expect that the interaction
between these spurious dipoles on \textbf{A} and \textbf{B}
might dominate the SAPT electrostatic contribution. If the
doubly occupied interfragment IBOs, together with the 
corresponding $+1$ nuclear charges, are instead reassigned to
fragments \textbf{A} and \textbf{B} or equally shared between
\textbf{A}/\textbf{B} and \textbf{C} \cite{Parrish:15}, the 
spurious dipole moments remain, and an additional problem 
appears when the linker is small and the resulting charge
densities of \textbf{A} and \textbf{B} are too close together.
For instance, in our propane example, such a reassignment is
not acceptable because it would lead to assigning a $+1$
nuclear charge {\em on the same central carbon atom} to
both \textbf{A} and \textbf{B}.

\begin{figure}[!h]
    \centering
    (a)\includegraphics[width=0.32\textwidth]{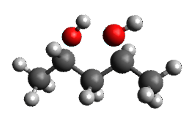}\hfil (b)\includegraphics[width=0.32\textwidth]{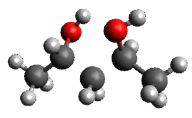}
    \caption{Two different partitionings of the 2,4-pentanediol molecule into hydrogen bonded fragments \textbf{A} and \textbf{B} covalently connected through a linking fragment \textbf{C}. A missing bond signifies a fragment boundary.}
    \label{fig:fragments}
\end{figure}

While the process of fragmentation and the definition of
intramolecular SAPT corrections is not unique as it does not
lead to any experimental observables, a useful approach
should lead to ISAPT results that make physical sense. 
For example, an intramolecular hydrogen bond should show up
as a favorable interaction both in terms of the total SAPT
energy and its electrostatic component, similar to an
intermolecular hydrogen bond. This is not always the case
for any of the variants (link orbital assignments) of the
original ISAPT method of Ref.~\citenum{Parrish:15}. A good
example is the 2,4-pentanediol molecule that involves an
intramolecular hydrogen bond. In the fragmentation scheme
of Fig.~\ref{fig:fragments} (a), which was extensively studied
in Ref.~\citenum{Parrish:15}, the ISAPT results make sense and
resemble the SAPT decomposition for the water dimer; in
particular, the electrostatic energy is attractive (negative).
However, in the fragmentation scheme of 
Fig.~\ref{fig:fragments} (b), the electrostatic energy is
strongly positive, which does not make physical sense as the
addition of nonpolar hydrocarbon chains to \textbf{A} and 
\textbf{B} cannot offset the favorable dipole-dipole 
interaction between the --OH groups. We believe that the
reason for this unphysical behavior are the artificial
dipole moments resulting from the interfragment boundaries
cutting through the (nonpolar) C-C bonds.

The problem of unphysical multipoles at the interfragment
boundary was already mentioned in Ref.~\citenum{Parrish:15},
and it was further explored by Meitei and Hesselmann in
Ref.~\citenum{Meitei:17}. These authors, in the specific
context of highly sterically crowded hydrocarbons such as
the all-meta tert-butyl derivative of hexaphenylethane
\cite{Schreiner:11}, proposed a solution to replace the
embedding in the linking fragment (in this case, two central
carbon atoms) by an embedding in a set of capping hydrogen
atoms at very short C-H bond lengths to reduce the artificial
dipoles. In most cases, this approach indeed led to a
sensible, somewhat attractive electrostatic energy; however,
such a correction is very system specific as it cannot be
expected that the embedding in the capping hydrogens can
replace an embedding in {\em any} linker fragment \textbf{C}.
Therefore, the issue of unphysical dipoles appearing at the
ISAPT interfragment boundary remains unsolved, limiting the 
applications of the method to highly favorable cases and/or 
cases where the errors cancel (which might be expected when 
investigating differences between similar systems).

In this work, we propose several algorithms to repartition the
link bonds between fragments to reduce the unphysical
fragment dipoles while preserving the embedding in the actual,
unaltered chemical system. In a sense, we follow a direction
stated, but not pursued, by Meitei and Hesselmann 
\cite{Meitei:17}, and explore an alternative assignment of 
the linking electron pair -- the one where one electron is
ascribed to each of the two adjacent fragments. This leads
to several new ISAPT variants, differing in the way in which
the linking IBO is deconstructed into a pair of intrinsic 
hybrid orbitals (IHOs) on the bonded atoms \cite{Knizia:13}.
It should be stressed that our goal is not the ultimate,
rigorous definition of ISAPT corrections (we are fully aware
that such fragment-fragment contributions are not 
measurable quantities), but a practical and robust ISAPT
algorithm whose results make physical sense for a wide
variety of molecular systems and fragmentation patterns.
In particular, we expect a practical ISAPT variant 
\begin{description}
\item[(a)] to substantially reduce the magnitude of the
fragment dipole moments when a nonpolar system is 
partitioned into nonpolar fragments,
\item[(b)] to give attractive electrostatic energies for 
intramolecular hydrogen
bonded fragments, and slightly attractive electrostatic energies
between nonpolar fragments (due to charge penetration 
\cite{Hohenstein:11b,Grynova:16}),
\item[(c)] to closely follow the results of standard
intermolecular SAPT at large \textbf{A}--\textbf{B} separations,
\item[(d)] to exhibit stable behavior as the one-electron basis
set is enlarged.
\end{description}

In the remainder of this article, we develop and test our
new ISAPT link bond partitionings with a goal of satisfying
the conditions \textbf{(a)}--\textbf{(d)} above. First, in
the Methodology section, we describe the new algorithms.
Next, the new partitionings are applied
to a set of representative intramolecular interactions and
compared with the original link assignment of 
Ref.~\citenum{Parrish:15}. The final section
contains conclusions.

\section{Methodology}\label{sec:method}

In the following discussion, the capital letters $K,L,M,\ldots$ will denote atomic orbital
(AO) basis functions (the AO basis always covers the entire molecule). The fragments \textbf{A} and
\textbf{B} interact with each other only noncovalently, but are covalently connected via the 
linker \textbf{C}. We will assume that (\textbf{A},\textbf{C}) and (\textbf{B},\textbf{C})
are each connected by a single bond, with $\phi_{AC}$ and $\phi_{BC}$ denoting the 
connecting IBOs \cite{Knizia:13}. The initial set of IBOs
is obtained by localizing the HF wavefunction for the
entire molecule, and its orthogonal subspaces containing
IBOs for the respective fragments will be denoted by
${\cal B}^0_{\mathbf A}$, ${\cal B}^0_{\mathbf B}$, and ${\cal B}_{\mathbf C}$.
Subsequently, the occupied orbitals of \textbf{A} and
\textbf{B} are reconstructed with the \textbf{A}--\textbf{B}
interaction switched off but the \textbf{A}--\textbf{C} and
\textbf{B}--\textbf{C} ones fully present.
Specifically, the HF equations for \textbf{A} and
\textbf{B} embedded in the frozen HF wavefunction of 
\textbf{C} are solved at this point.
Afterwards, the entire one-electron space
${\cal B}$ (the space spanned by the AOs) includes occupied subspaces 
${\cal B}_{\mathbf A}$, ${\cal B}_{\mathbf B}$, ${\cal B}_{\mathbf C}$ for the respective
fragments, with both ${\cal B}_{\mathbf A}$ and ${\cal B}_{\mathbf B}$ orthogonal to
${\cal B}_{\mathbf C}$ (however, ${\cal B}_{\mathbf A}$ is
not orthogonal to ${\cal B}_{\mathbf B}$; note also that the
subspace for the linker \textbf{C} has not changed in the
last step). The virtual orbitals for fragments \textbf{A} and \textbf{B} live
in the orthogonal complements to the spaces ${\cal B}_{\mathbf A}\oplus{\cal B}_{\mathbf C}$ 
and ${\cal B}_{\mathbf B}\oplus{\cal B}_{\mathbf C}$, respectively. Thus, the occupied orbitals
on \textbf{C} are projected out from both the occupied and virtual orbitals on 
\textbf{A}/\textbf{B}.

In the original ISAPT method
\cite{Parrish:15} with the default link assignment, the doubly occupied orbitals $\phi_{AC}$ 
and $\phi_{BC}$ are assigned to linker \textbf{C} together with a single proton from the
nucleus of \textbf{A}/\textbf{B} that participates in the covalent bond (the remaining protons
are still assigned to \textbf{A}/\textbf{B}, so that this nucleus has an effective charge reduced
by one within that fragment). Thus, if the original molecule was electrically neutral, all 
fragments are neutral too.
Unfortunately, the same cannot be said 
about the dipole moments on the individual fragments.
The connecting atom on \textbf{A} or \textbf{B}, besides missing
one proton, is missing an electron on one of its hybrid orbitals, the one responsible for the
covalent bond with \textbf{C}. Thus, even for a nonpolar environment such as an sp$^3$ carbon
in a hydrocarbon molecule, an artificial dipole moment is created on the fragment 
\textbf{A}/\textbf{B} because one hybrid orbital, with its bonding electron, has been
removed. We believe that this artificial dipole moment is the cause of often
unphysical interfragment electrostatic energies in ISAPT.

The IBOs $\phi_{AC}$ and $\phi_{BC}$ can be viewed as linear combinations of ``intrinsic hybrid
orbitals'' (IHOs) on the atoms connected by the linking bond --- combinations of intrinsic atomic orbitals (IAOs) \cite{Knizia:13} resulting 
in a hybrid orbital pointing in the bond direction. We propose to partially undo the 
IAO$\to$IBO transformation and identify IHOs $\chi_x$ and $\chi_y$ that, in some sense, 
constitute the \textbf{A}-fragment and \textbf{B}-fragment parts of $\phi_{AC}$ and $\phi_{BC}$,
respectively. 
More than one algorithm of determining these IHOs is possible and we will present two specific
choices below. Normally, we do assume that $\chi_x$ and $\chi_y$, like the whole link orbital 
space where they come from, are orthogonal to the occupied spaces ${\cal B}_{\mathbf A}$ and 
${\cal B}_{\mathbf B}$, respectively. Note that this orthogonalization will have to be 
explicitly enforced as $\chi_x$ and $\chi_y$ are not contained in the
subspace ${\cal B}_{\mathbf C}$; we will also investigate
what happens if one forgoes this orthogonalization.
In this process of bond orbital decomposition, we associate one electron with each of
$\chi_x$ and $\chi_y$ and, for the purpose of updating density matrices, we will
assume that half of this electron comes in with spin up and half with spin down. In our new
link assignments, the single electron on $\chi_x$ is reassigned from \textbf{C} to
\textbf{A} together with the corresponding +1 nuclear charge on the \textbf{A} atom connected
to \textbf{C}. Thus, the assignment of nuclear charges is now very simple -- the full charges of 
all nuclei in \textbf{A} belong to \textbf{A} -- but the assignment of electronic charges becomes
more complicated. An analogous reassignment takes place between \textbf{C} and \textbf{B} ---
the net result is that \textbf{A} and \textbf{B} gain one electron each relative to the default
ISAPT link assignment. Our hope is that, by the addition of electrons on 
$\chi_x$ and $\chi_y$, the fragment charges around the linking atoms become more spherically
symmetrical, and any unphysical dipole moments at the fragment boundary should be substantially
diminished.

At this point, the redefinition of the ISAPT first-order electrostatic energy
is straightforward. The nuclear potentials of fragments  \textbf{A} and \textbf{B} are modified
to account for the entire charge $+Ze^{-}$, not $+(Z-1)e^{-}$, on the linking atom. The
HF-level fragment density matrices $D_{\mathbf A}$ and $D_{\mathbf B}$ are supplemented by the
contributions of the (singly occupied) orbitals $\chi_x$ and $\chi_y$, respectively.
The updated
nuclear potentials and density matrices are then used in standard SAPT expressions to 
determine the ISAPT $E^{(10)}_{\rm elst}$ term between
fragments \textbf{A} and \textbf{B}. 

The $E^{(10)}_{\rm exch}$ correction, within its customary $S^2$ approximation \cite{Jeziorski:94}, requires, in addition to the full density matrix, its partitioning into spin density matrices.
It is natural to assume that the
supplementary contributions of $\chi_x$ and $\chi_y$ are divided equally between the spin-up and spin-down density
matrix (half spin-up and half spin-down). 
However, this assumption does not determine the exchange energy uniquely, as the interactions between the electrons on $\chi_x$ and $\chi_y$ depend on the mutual alignment of their spins.
In the two limiting cases that both lead to equal spin-up and spin-down density matrices, the link spinorbitals $\psi_x$ and $\psi_y$ can have parallel spins,
\begin{equation}
\psi_x^{\parallel} = \frac{1}{\sqrt{2}}\chi_x \left(|\uparrow\rangle+|\downarrow\rangle\right) \;\;\;\;\;\;
\psi_y^{\parallel} = \frac{1}{\sqrt{2}}\chi_y \left(|\uparrow\rangle+|\downarrow\rangle\right)
\label{eq:psixypar}    
\end{equation}
or perpendicular spins,
\begin{equation}
\psi_x^{\perp} = \frac{1}{\sqrt{2}}\chi_x \left(|\uparrow\rangle+|\downarrow\rangle\right) \;\;\;\;\;\;
\psi_y^{\perp} = \frac{1}{\sqrt{2}}\chi_y \left(|\uparrow\rangle-|\downarrow\rangle\right)
\label{eq:psixyperp}    
\end{equation}
Note that the parallel case corresponds to $\langle \psi_x^{\parallel}|\psi_y^{\parallel}\rangle=\langle \chi_x|\chi_y\rangle$ and includes one electron worth of exchange interaction between the link spinorbitals.
The perpendicular case leads to $\langle \psi_x^{\perp}|\psi_y^{\perp}\rangle=0$; thus, no exchange interaction between the link spinorbitals exists.
Both cases introduce an interdependence between the spins of the \textbf{A--C} and \textbf{B--C} link electrons which can be viewed as unphysical --- the \textbf{A--C} and \textbf{B--C} partitionings should be independent of each other, leading to a random mutual orientation of spins.
To model this random spin coupling, we propose to adopt as the final exchange energy the average of the parallel and perpendicular approaches:
\begin{equation}
E^{(10)}_{\rm exch}(S^2) = \frac{1}{2}\left(E^{(10)\parallel}_{\rm exch}(S^2)+ E^{(10)\perp}_{\rm exch}(S^2)\right)
\label{eq:e10xavg}    
\end{equation}
Following the standard derivation of MO-based exchange corrections within the density-matrix formalism \cite{Moszynski:94a} (valid in both dimer and monomer basis sets) and recasting the resulting formulas to the AO basis \cite{Hesselmann:05,Patkowski:18,Lao:18}, we obtain the following formulas for the parallel-spin and perpendicular-spin variants:
\begin{align}
E^{(10)\parallel/\perp}_{\rm exch}(S^2)=&
-2\mathbf{D}^A\cdot\mathbf{K}^B
\mp \frac{1}{2}\mathbf{D}^X\cdot\mathbf{K}^Y
-2\mathbf{V}^A\cdot (\mathbf{D}^A\mathbf{S}^{AO}\mathbf{D}^B)
\mp\frac{1}{2} \mathbf{V}^A\cdot (\mathbf{D}^X\mathbf{S}^{AO}\mathbf{D}^Y)
\nonumber \\ &
-4\mathbf{J}^A\cdot (\mathbf{D}^A\mathbf{S}^{AO}\mathbf{D}^B)
\mp \mathbf{J}^A\cdot (\mathbf{D}^X\mathbf{S}^{AO}\mathbf{D}^Y)
+2\mathbf{K}^A\cdot (\mathbf{D}^A\mathbf{S}^{AO}\mathbf{D}^B)
\nonumber \\ &
+\frac{1}{2}\mathbf{K}^X\cdot (\mathbf{D}^X\mathbf{S}^{AO}\mathbf{D}^B)
\pm \frac{1}{2}\mathbf{K}^X\cdot (\mathbf{D}^A\mathbf{S}^{AO}\mathbf{D}^Y)
\pm \frac{1}{2}\mathbf{K}^A\cdot (\mathbf{D}^X\mathbf{S}^{AO}\mathbf{D}^Y)
\nonumber \\ &
-2\mathbf{V}^B\cdot (\mathbf{D}^B\mathbf{S}^{AO}\mathbf{D}^A)
\mp\frac{1}{2} \mathbf{V}^B\cdot (\mathbf{D}^Y\mathbf{S}^{AO}\mathbf{D}^X)
\nonumber \\ &
-4\mathbf{J}^B\cdot (\mathbf{D}^B\mathbf{S}^{AO}\mathbf{D}^A)
\mp \mathbf{J}^B\cdot (\mathbf{D}^Y\mathbf{S}^{AO}\mathbf{D}^X)
+2\mathbf{K}^B\cdot (\mathbf{D}^B\mathbf{S}^{AO}\mathbf{D}^A)
\nonumber \\ &
+\frac{1}{2}\mathbf{K}^Y\cdot (\mathbf{D}^Y\mathbf{S}^{AO}\mathbf{D}^A)
\pm \frac{1}{2}\mathbf{K}^Y\cdot (\mathbf{D}^B\mathbf{S}^{AO}\mathbf{D}^X)
\pm \frac{1}{2}\mathbf{K}^B\cdot (\mathbf{D}^Y\mathbf{S}^{AO}\mathbf{D}^X)
\nonumber \\ &
+2\mathbf{V}^A\cdot (\mathbf{D}^B\mathbf{S}^{AO}\mathbf{D}^A\mathbf{S}^{AO}\mathbf{D}^B)
+\frac{1}{2}\mathbf{V}^A\cdot (\mathbf{D}^Y\mathbf{S}^{AO}\mathbf{D}^A\mathbf{S}^{AO}\mathbf{D}^Y)
\nonumber \\ &
\pm\frac{1}{2} \mathbf{V}^A\cdot (\mathbf{D}^Y\mathbf{S}^{AO}\mathbf{D}^X\mathbf{S}^{AO}\mathbf{D}^B)
\pm\frac{1}{2} \mathbf{V}^A\cdot (\mathbf{D}^B\mathbf{S}^{AO}\mathbf{D}^X\mathbf{S}^{AO}\mathbf{D}^Y)
\nonumber \\ &
+4\mathbf{J}^A\cdot (\mathbf{D}^B\mathbf{S}^{AO}\mathbf{D}^A\mathbf{S}^{AO}\mathbf{D}^B)
+\mathbf{J}^A\cdot (\mathbf{D}^Y\mathbf{S}^{AO}\mathbf{D}^A\mathbf{S}^{AO}\mathbf{D}^Y)
\nonumber \\ &
\pm \mathbf{J}^A\cdot (\mathbf{D}^Y\mathbf{S}^{AO}\mathbf{D}^X\mathbf{S}^{AO}\mathbf{D}^B)
\pm \mathbf{J}^A\cdot (\mathbf{D}^B\mathbf{S}^{AO}\mathbf{D}^X\mathbf{S}^{AO}\mathbf{D}^Y)
\nonumber \\ &
+2\mathbf{V}^B\cdot (\mathbf{D}^A\mathbf{S}^{AO}\mathbf{D}^B\mathbf{S}^{AO}\mathbf{D}^A)
+\frac{1}{2}\mathbf{V}^B\cdot (\mathbf{D}^X\mathbf{S}^{AO}\mathbf{D}^B\mathbf{S}^{AO}\mathbf{D}^X)
\nonumber \\ &
\pm\frac{1}{2} \mathbf{V}^B\cdot (\mathbf{D}^X\mathbf{S}^{AO}\mathbf{D}^Y\mathbf{S}^{AO}\mathbf{D}^A)
\pm\frac{1}{2} \mathbf{V}^B\cdot (\mathbf{D}^A\mathbf{S}^{AO}\mathbf{D}^Y\mathbf{S}^{AO}\mathbf{D}^X)
\nonumber \\ &
+4\mathbf{J}^B\cdot (\mathbf{D}^A\mathbf{S}^{AO}\mathbf{D}^B\mathbf{S}^{AO}\mathbf{D}^A)
+\mathbf{J}^B\cdot (\mathbf{D}^X\mathbf{S}^{AO}\mathbf{D}^B\mathbf{S}^{AO}\mathbf{D}^X)
\nonumber \\ &
\pm \mathbf{J}^B\cdot (\mathbf{D}^X\mathbf{S}^{AO}\mathbf{D}^Y\mathbf{S}^{AO}\mathbf{D}^A)
\pm \mathbf{J}^B\cdot (\mathbf{D}^A\mathbf{S}^{AO}\mathbf{D}^Y\mathbf{S}^{AO}\mathbf{D}^X)
\nonumber \\ &
-2 (\mathbf{D}^A\mathbf{S}^{AO}\mathbf{D}^B)\cdot
\mathbf{K}^{\dagger}[\mathbf{D}^A\mathbf{S}^{AO}\mathbf{D}^B]
-\frac{1}{2}(\mathbf{D}^X\mathbf{S}^{AO}\mathbf{D}^B)\cdot
\mathbf{K}^{\dagger}[\mathbf{D}^X\mathbf{S}^{AO}\mathbf{D}^B]
\nonumber \\ &
-\frac{1}{2}(\mathbf{D}^A\mathbf{S}^{AO}\mathbf{D}^Y)\cdot
\mathbf{K}^{\dagger}[\mathbf{D}^A\mathbf{S}^{AO}\mathbf{D}^Y]
\mp\frac{1}{2}(\mathbf{D}^A\mathbf{S}^{AO}\mathbf{D}^Y)\cdot
\mathbf{K}^{\dagger}[\mathbf{D}^X\mathbf{S}^{AO}\mathbf{D}^B]
\nonumber \\ &
\mp\frac{1}{2}(\mathbf{D}^A\mathbf{S}^{AO}\mathbf{D}^B)\cdot
\mathbf{K}^{\dagger}[\mathbf{D}^X\mathbf{S}^{AO}\mathbf{D}^Y]
\mp\frac{1}{2}(\mathbf{D}^X\mathbf{S}^{AO}\mathbf{D}^Y)\cdot
\mathbf{K}^{\dagger}[\mathbf{D}^A\mathbf{S}^{AO}\mathbf{D}^B]
\nonumber \\ &
\mp\frac{1}{2}(\mathbf{D}^X\mathbf{S}^{AO}\mathbf{D}^B)\cdot
\mathbf{K}^{\dagger}[\mathbf{D}^A\mathbf{S}^{AO}\mathbf{D}^Y]
-\frac{1}{8}(\mathbf{D}^X\mathbf{S}^{AO}\mathbf{D}^Y)\cdot
\mathbf{K}^{\dagger}[\mathbf{D}^X\mathbf{S}^{AO}\mathbf{D}^Y]
\label{e10s2:parper}
\end{align}
In the above equation, the upper plus/minus signs correspond to $E^{(10)\parallel}_{\rm exch}(S^2)$ and the lower signs to $E^{(10)\perp}_{\rm exch}(S^2)$. 
Furthermore, the AO-basis density matrices are computed from the LCAO MO coefficients $C_{iK}$ in the usual way: 
\begin{align}
&(\mathbf{D}^A)_{KL}=\sum_i C_{iK}C_{iL}+\frac{1}{2}C_{xK}C_{xL} \;\;\;\;\;\;
(\mathbf{D}^B)_{KL}=\sum_j C_{jK}C_{jL}+\frac{1}{2}C_{yK}C_{yL} \nonumber \\ &
(\mathbf{D}^X)_{KL}=C_{xK}C_{xL} \;\;\;\;\;\;
(\mathbf{D}^Y)_{KL}=C_{yK}C_{yL}
\label{eq:dadb}    
\end{align}
where the indices $i$ and $j$ run over occupied orbitals of $\mathbf{A}$ and $\mathbf{B}$, respectively, and indices $x,y$ correspond to the link orbitals $\chi_x,\chi_y$ (the factors $\frac{1}{2}$ result from the link orbital containing only half an electron of a given spin).
Note that $\mathbf{D}^A$ and $\mathbf{D}^B$ already contain the link-orbital contributions $\mathbf{D}^X$ and $\mathbf{D}^Y$, respectively, but the numerical factors multiplying the link-orbital terms need to be adjusted by the presence of the additional terms involving $\mathbf{D}^X$ and $\mathbf{D}^Y$. 
The elements of the commonly used generalized Coulomb and exchange matrices are given by
\begin{equation}
\mathbf{J}[\mathbf{X}]_{KL}=\sum_{MN}(KL|MN)\mathbf{X}_{MN}
\;\;\;\;\;\;
\mathbf{K}[\mathbf{X}]_{KL}=\sum_{MN}(KM|NL)\mathbf{X}_{MN}
\label{eq:JKmat}    
\end{equation}
the shorthand notations $\mathbf{J}^{Z}\equiv\mathbf{J}[\mathbf{D}^{Z}]$ and $\mathbf{K}^{Z}\equiv\mathbf{K}[\mathbf{D}^{Z}]$ $(Z=A,B,X,Y)$ indicate the regular (not generalized) Coulomb and exchange matrices, $\mathbf{V}^A/\mathbf{V}^B$ are matrices containing the nuclear attraction integrals for an appropriate fragment, $\mathbf{S}^{AO}$ is the AO-basis overlap matrix $(\mathbf{S}^{AO})_{KL}=\langle K|L\rangle$,
and the dot signifies an inner product of matrices:
\begin{equation}
\mathbf{X}\cdot\mathbf{Y}=\sum_{KL}\mathbf{X}_{KL}\mathbf{Y}_{KL}
\label{eq:innerprod}
\end{equation}
The case of the full nonapproximated $E^{(10)}_{\rm exch}$ is more complicated and
will be presented in detail in Appendix A.

A more subtle issue is defining a suitable induction energy 
$E^{(20)}_{\rm Ind}$ (following Ref.~\citenum{Schaffer:12}, 
the capitalized correction name denotes a sum of pure 
induction and exchange-induction terms) and, in a 
coherent manner, the $\delta_{\rm HF}$ correction that captures higher-order induction and
exchange-induction effects beyond that specific form of $E^{(20)}_{\rm Ind}$. For these purposes,
we note that the single electron on $\chi_x$ ($\chi_y$) does contribute (via the
electron density) to the electrostatic potential of a given fragment that polarizes the other
one. However, the orbitals $\chi_x$ ($\chi_y$) themselves are assumed to be frozen
(unpolarizable). In other words, we consider the response (polarization) of the occupied 
orbitals in ${\cal B}_{\mathbf A}$ under the influence of the electrostatic potential resulting
from the nuclei of \textbf{B} and the occupied orbitals in 
${\cal B}_{\mathbf B}\oplus (1/2) \{\chi_y\}$ and vice versa, where ``(1/2)'' reminds that
the orbital $\chi_y$ is singly occupied while the orbitals spanning ${\cal B}_{\mathbf B}$
are doubly occupied. 
A technical issue is the precise choice of the virtual space for the
response of fragment-\textbf{A} orbitals. We employed for this virtual space the
orthogonal complement of ${\cal B}_{\mathbf A}\oplus{\cal B}_{\mathbf C}$ like in the default ISAPT variant,
disregarding the fact that the IHO $\chi_x$, carved out of ${\cal B}_{\mathbf C}$, is not precisely contained in ${\cal B}_{\mathbf C}$ and has a nonzero component in the unoccupied space.

The coupled perturbed Hartree-Fock (CPHF) coefficients $C^i_a/C^j_b$ for each fragment and the pure induction correction $E^{(20)}_{\rm ind,resp}$ can now be computed using standard SAPT0 formulas (note that $a$ and $b$ denote virtual orbitals of \textbf{A} and \textbf{B}, respectively).
The exchange-induction correction $E^{(20)}_{\rm exch-ind,resp}$, computed in ISAPT within the $S^2$ approximation, again depends on whether a parallel (Eq.~(\ref{eq:psixypar})) or perpendicular (Eq.~(\ref{eq:psixyperp})) spin coupling of link electrons is assumed, and we will again opt for the average of the two as our final correction. 
The resulting AO-based formula for both couplings, derived via the density matrix formalism \cite{Moszynski:94a}, reads
\begin{align}
E^{(20)\parallel/\perp}_{\rm exch-ind,resp}({\rm A}\leftarrow{\rm B}, S^2)=&\mathbf{C}^{A}\cdot
\left[-2\mathbf{K}^B
-4\mathbf{J}^B\mathbf{D}^B\mathbf{S}^{AO}
-2\mathbf{V}^B\mathbf{D}^B\mathbf{S}^{AO}
+2\mathbf{K}^B\mathbf{D}^B\mathbf{S}^{AO}
\right. \nonumber \\ &
+\frac{1}{2}\mathbf{K}^Y\mathbf{D}^Y\mathbf{S}^{AO}
-4\mathbf{J}[\mathbf{D}^A\mathbf{S}^{AO}\mathbf{D}^B]
\mp \mathbf{J}[\mathbf{D}^X\mathbf{S}^{AO}\mathbf{D}^Y]
\nonumber \\ &
+2\mathbf{S}^{AO}\mathbf{D}^B\mathbf{K}^A
\pm\frac{1}{2}\mathbf{S}^{AO}\mathbf{D}^Y\mathbf{K}^X
+2\mathbf{K}[\mathbf{D}^A\mathbf{S}^{AO}\mathbf{D}^B]
\nonumber \\ &
\pm\frac{1}{2}\mathbf{K}[\mathbf{D}^X\mathbf{S}^{AO}\mathbf{D}^Y]
-4\mathbf{S}^{AO}\mathbf{D}^B\mathbf{J}^A
-2\mathbf{S}^{AO}\mathbf{D}^B\mathbf{V}^A
\nonumber \\ &
+4\mathbf{S}^{AO}\mathbf{D}^B\mathbf{S}^{AO}\mathbf{D}^A\mathbf{J}^B
\pm \mathbf{S}^{AO}\mathbf{D}^Y\mathbf{S}^{AO}\mathbf{D}^X\mathbf{J}^B
\nonumber \\ &
+2\mathbf{S}^{AO}\mathbf{D}^B\mathbf{S}^{AO}\mathbf{D}^A\mathbf{V}^B
\pm\frac{1}{2} \mathbf{S}^{AO}\mathbf{D}^Y\mathbf{S}^{AO}\mathbf{D}^X\mathbf{V}^B
\nonumber \\ &
+4\mathbf{J}^B\mathbf{D}^A\mathbf{S}^{AO}\mathbf{D}^B\mathbf{S}^{AO}
\pm \mathbf{J}^B\mathbf{D}^X\mathbf{S}^{AO}\mathbf{D}^Y\mathbf{S}^{AO}
\nonumber \\ &
+2\mathbf{V}^B\mathbf{D}^A\mathbf{S}^{AO}\mathbf{D}^B\mathbf{S}^{AO}
\pm\frac{1}{2}\mathbf{V}^B\mathbf{D}^X\mathbf{S}^{AO}\mathbf{D}^Y\mathbf{S}^{AO}
\nonumber \\ &
+4\mathbf{J}[\mathbf{D}^B\mathbf{S}^{AO}\mathbf{D}^A\mathbf{S}^{AO}\mathbf{D}^B]
\pm\mathbf{J}[\mathbf{D}^Y\mathbf{S}^{AO}\mathbf{D}^X\mathbf{S}^{AO}\mathbf{D}^B]
\nonumber \\ &
\pm\mathbf{J}[\mathbf{D}^B\mathbf{S}^{AO}\mathbf{D}^X\mathbf{S}^{AO}\mathbf{D}^Y]
+\mathbf{J}[\mathbf{D}^Y\mathbf{S}^{AO}\mathbf{D}^A\mathbf{S}^{AO}\mathbf{D}^Y]
\nonumber \\ &
-2\mathbf{S}^{AO}\mathbf{D}^B\mathbf{K}[\mathbf{D}^B\mathbf{S}^{AO}\mathbf{D}^A]
\mp\frac{1}{2}\mathbf{S}^{AO}\mathbf{D}^B\mathbf{K}[\mathbf{D}^Y\mathbf{S}^{AO}\mathbf{D}^X]
\nonumber \\ &
\mp\frac{1}{2}\mathbf{S}^{AO}\mathbf{D}^Y\mathbf{K}[\mathbf{D}^B\mathbf{S}^{AO}\mathbf{D}^X]
-\frac{1}{2}\mathbf{S}^{AO}\mathbf{D}^Y\mathbf{K}[\mathbf{D}^Y\mathbf{S}^{AO}\mathbf{D}^A]
\nonumber \\ &
-2\mathbf{K}[\mathbf{D}^A\mathbf{S}^{AO}\mathbf{D}^B]\mathbf{D}^B\mathbf{S}^{AO}
\mp\frac{1}{2}\mathbf{K}[\mathbf{D}^X\mathbf{S}^{AO}\mathbf{D}^B]\mathbf{D}^Y\mathbf{S}^{AO}
\nonumber \\ &
\mp\frac{1}{2}\mathbf{K}[\mathbf{D}^X\mathbf{S}^{AO}\mathbf{D}^Y]\mathbf{D}^B\mathbf{S}^{AO}
-\frac{1}{2}\mathbf{K}[\mathbf{D}^A\mathbf{S}^{AO}\mathbf{D}^Y]\mathbf{D}^Y\mathbf{S}^{AO}
\nonumber \\ &
+4\mathbf{S}^{AO}\mathbf{D}^B\mathbf{J}^A\mathbf{D}^B\mathbf{S}^{AO}
+\mathbf{S}^{AO}\mathbf{D}^Y\mathbf{J}^A\mathbf{D}^Y\mathbf{S}^{AO}
\nonumber \\ & \left.
+2\mathbf{S}^{AO}\mathbf{D}^B\mathbf{V}^A\mathbf{D}^B\mathbf{S}^{AO}
+\frac{1}{2}\mathbf{S}^{AO}\mathbf{D}^Y\mathbf{V}^A\mathbf{D}^Y\mathbf{S}^{AO}
\right]
\label{eq:e20xiparper}
\end{align}
where $\mathbf{C}^A$ is the matrix of fragment-A CPHF coefficients $C^i_a$ backtransformed to the AO basis using the appropriate LCAO MO coefficients $C_{iK}$ and $C_{aL}$ \cite{Smith:18}:
\begin{equation}
\mathbf{C}^A_{KL}=C^i_a C_{iK}C_{aL}
\label{eq:CPHFao}    
\end{equation}
An analogous formula for the other exchange induction term $E^{(20)\parallel/\perp}_{\rm exch-ind,resp}({\rm B}\leftarrow{\rm A}, S^2)$
(where, this time, fragment $\mathbf{A}$ polarizes fragment $\mathbf{B}$)
is obtained from Eq.~(\ref{eq:e20xiparper}) by an exchange of all symbols pertaining to $\mathbf{A}$ by the corresponding ones of $\mathbf{B}$ and vice versa.

The adjustment to the treatment of second-order exchange-dispersion energy resulting from our link bond reassignment is quite analogous to exchange-induction energy (note that the $E^{(20)}_{\rm disp}$ term needs no adjustment as excitations from the link orbital are not considered). 
The resulting AO-based formulas for $E^{(20)\parallel/\perp}_{\rm exch-disp}(S^2)$ are presented in the Supporting Information.
Notably, all additional terms resulting from the link-electron spin coupling cancel out when averaging the parallel and perpendicular cases, so the spin-averaged value of $E^{(20)}_{\rm exch-disp}(S^2)$ is given by the standard SAPT0 formula (see e.g. Ref.~\citenum{Smith:18} for its AO form) without any modifications.

Unfortunately, our proposed definition of interfragment
induction energy does not lend itself to an infinite-order
generalization that can be used to define $\delta_{\rm HF}$.
On the noninteracting fragment side, the induction energy accounts for a polarization of orbitals from ${\cal B}_{\mathbf A}$
in the electrostatic potential of ${\cal B}_{\mathbf B}\oplus (1/2) \{\chi_y\}$ and the
other way around. 
Accordingly, one should optimize the noninteracting occupied orbitals for \textbf{A} in the orthogonal complement of 
${\cal B}_{\mathbf C}$, with the embedding
potential including interactions with nuclei of \textbf{C} and the sum of electron densities
coming from $\chi_x$ and the rest of ${\cal B}_{\mathbf C}$ (recall that the latter has had two
electrons removed and reassigned to \textbf{A} and \textbf{B}). Note that the embedding potential
does not contain the interaction with the single-electron density from $\chi_y$, the IHO at the
more distant fragment boundary, which has been reassigned
to fragment \textbf{B}; thus, the entire 
\textbf{A}--\textbf{B} interaction has been temporarily switched off.

On the interacting side, one optimizes the occupied orbitals of 
\textbf{AB} also in the orthogonal
complement of ${\cal B}_{\mathbf C}$, with the nuclei of
\textbf{C} and the Coulomb and exchange operators corresponding to the density
of ${\cal B}_{\mathbf C}$ forming the HF embedding potential.
Note that this potential is obtained from the same HF calculation for the entire molecule.
Moreover, the
reassignment of electrons on $\chi_x$ and $\chi_y$ does not matter this time because the HF system
is embedded in the sum of densities coming from $\chi_x$, $\chi_y$, and the rest of 
${\cal B}_{\mathbf C}$. The orbitals $\chi_x$ and $\chi_y$, while frozen in the
HF optimization, do belong to the fragments \textbf{A} and \textbf{B}, so that their
interaction with each other and with the nuclei of the other fragment needs to be included in the HF interaction energy
(it is a constant term that does not depend on the orbitals being optimized). 
Unfortunately, in this algorithm, the treatment of the two
electrons assigned to $\chi_x$ and $\chi_y$ is not consistent
between the fragment calculations, where they are described
by frozen orbitals, and the molecular one, where they belong
to doubly occupied orbitals which are variationally
optimized. As a result, we were not able to define a 
supermolecular $\delta_{\rm HF}$ correction that is consistent
with the proposed reassignment of link bond electrons and
the resulting $E^{(10)}_{\rm elst}$, $E^{(10)}_{\rm exch}$,
$E^{(20)}_{\rm ind,resp}$, and $E^{(20)}_{\rm exch-ind,resp}$
corrections. In the calculations below, this term will be
taken from ISAPT calculations with the default assignment
of the entire link orbitals to \textbf{C}.

We note at this point that the embedding defined in the fragment HF calculations described above
(omitting the IHO at the distant fragment boundary), while perfectly consistent with
our electrostatic and induction energy expressions, is not the same as the embedding used to compute the 
initial HF orbitals spanning ${\cal B}_{\mathbf A}$ and ${\cal B}_{\mathbf B}$ (which, by
default, always takes into account the entire linker including both $\chi_x$ and $\chi_y$).
If the linker is sufficiently large, the difference in the embedding density occurs far away
from the fragment in question and its effect should be small, however, we have to consider
whether an orbital reoptimization is in order before the computation of ISAPT corrections. 
On the other hand, $\chi_x$ and $\chi_y$ depend on the spaces ${\cal B}_{\mathbf A}$ and ${\cal B}_{\mathbf B}$
because they are orthogonalized to these spaces.
This circular dependence suggests that iteration to self-consistency is the most
rigorous way to implement the process: determine $\chi_x$ and $\chi_y$ which are orthogonal to the initial spaces ${\cal B}_{\mathbf A}$ and ${\cal B}_{\mathbf B}$ 
from the default link assignment algorithm, reoptimize ${\cal B}_{\mathbf A}$ in an embedding excluding $\chi_y$ and ${\cal B}_{\mathbf B}$ in an embedding excluding $\chi_x$, update $\chi_x/\chi_y$ by enforcing
orthogonality to new ${\cal B}_{\mathbf A}/{\cal B}_{\mathbf B}$, and so on. 
However, a fully self-consistent algorithm is impractical --- each iteration requires converged HF calculations for both 
fragments. On the other hand, as differences in embedding 
should be minor and occur relatively far away from a given fragment, this process should be very quickly convergent, 
and we can limit ourselves to zero, one, or two iterations. This leads to three variants of our approach:
\begin{description}
\item[Zero iterations:] the occupied spaces ${\cal B}_{\mathbf A}/{\cal B}_{\mathbf B}$ are taken directly from the default ISAPT variant, that is, the HF orbitals have been obtained using an embedding in the entire ${\cal B}_{\mathbf C}$ space. These occupied spaces are used to construct $\chi_x$ and $\chi_y$, and the ISAPT corrections are obtained from the resulting augmented density matrices and the polarization of the occupied spaces. 
\item[One iteration:] after the IHOs $\chi_x$ and $\chi_y$ are constructed as above, the HF calculations for fragments are repeated, this time with the embedding potential for \textbf{A} excluding $\chi_y$ and the embedding potential for \textbf{B} excluding $\chi_x$. This leads to new occupied spaces ${\cal B}'_{\mathbf A}/{\cal B}'_{\mathbf B}$, which in turn results in updated IHOs $\chi'_x$ and $\chi'_y$ orthogonal to the new occupied spaces. Now, all ISAPT corrections are obtained from the resulting augmented density matrices and the polarization of the occupied spaces ${\cal B}'_{\mathbf A}/{\cal B}'_{\mathbf B}$.
\item[Two iterations:] after the IHOs $\chi'_x$ and $\chi'_y$ are constructed as above, the HF calculations for fragments are repeated once again, this time with the embedding potential for \textbf{A} excluding $\chi'_y$ and the embedding potential for \textbf{B} excluding $\chi'_x$. This leads to new occupied spaces ${\cal B}''_{\mathbf A}/{\cal B}''_{\mathbf B}$, which in turn results in updated IHOs $\chi''_x$ and $\chi''_y$ orthogonal to the new occupied spaces. Now, all ISAPT corrections are obtained as above, but using the occupied spaces ${\cal B}''_{\mathbf A}/{\cal B}''_{\mathbf B}$ and the IHOs
$\chi''_x$ and $\chi''_y$.
\end{description}

To complete the specification of our method, we have to define how the IHOs $\chi_x$ and $\chi_y$ are determined. 
We propose two algorithms for this, both
of which proceed with a projection step (onto the fragment
\textbf{A} or \textbf{B}) and an orthogonalization step
(relative to all doubly occupied orbitals of this fragment).
Once the specific IBO $\phi_{AC}$ responsible for the \textbf{A}--\textbf{C} link bond is identified, its
projection onto \textbf{A} can be performed in one of two
ways. In the first approach, inspired by the ALMO method \cite{Khaliullin:07}, 
$\phi_{AC}$ is represented in the AO basis and
projected onto the space of basis functions centered on atoms of \textbf{A} only (all other coefficients of
this orbital are zeroed). This approach will be termed
Splitting of Atomic Orbitals (SAO).
In the second approach, $\phi_{AC}$ is represented in the IAO basis \cite{Knizia:13}, projected onto the space of IAOs
centered on atoms of \textbf{A} only by zeroing all other
coefficients, and transformed back to the AO basis.
This approach will be termed
Splitting of Intrinsic Atomic Orbitals (SIAO).
In either method, the final IHO $\chi_x$ is obtained from
the projected orbital by a Schmidt orthonormalization to all 
doubly occupied orbitals of \textbf{A}, that is, to the ${\cal B}_{\mathbf A}$ space.
Note that the SIAO projection is not strictly an ALMO 
(it has nonzero coefficients of basis functions on the other fragments resulting from the tails of the IAOs), and the
SAO projection ceases to be an ALMO after
orthogonalization. However, in either case, $\chi_x$ is still 
mostly localized on \textbf{A}.
The same projection-then-orthogonalization procedure is 
performed at the other interfragment boundary to determine
$\chi_y$. 

The final specification of the new ISAPT 
algorithm includes both the SAO/SIAO recipe to determine
link IHOs and the level of self-consistency between those 
IHOs and the occupied orbital space expressed by the number
of iterations described earlier in this section. For example,
the SAO0 method involves the SAO projection algorithm
resulting in $\chi_x/\chi_y$ orthogonalized to the
original occupied spaces ${\cal B}_{\mathbf A}/{\cal B}_{\mathbf B}$ (zero iterations), 
while the SIAO2 one uses the SIAO projection leading to
$\chi''_x/\chi''_y$ orthogonalized to the
twice updated occupied spaces ${\cal B}''_{\mathbf A}/{\cal B}''_{\mathbf B}$.
While the orthogonalization step is formally required
to obtain a valid fragment density matrix when augmented by
$\chi_x/\chi_y$, we will perform a limited set of numerical
tests to investigate the practical consequences of not
orthogonalizing (note that the lack of orthogonalization 
makes the algorithms with 1 and 2 iterations identical as
$\chi_X\equiv\chi'_X\equiv\chi''_X$ for $X=x,y$).

Finally, we note a common formal flaw of both selections
of the IHOs $\chi_x$ and $\chi_y$: they are not entirely
contained in the occupied space ${\cal B}_{\mathbf C}$.
As a result, from the point of view of the linker \textbf{C},
the reassignment of one electron each to $\chi_x$ and $\chi_y$
involves subtracting some electron density that is not
present in the system, leaving out a residual density matrix
for \textbf{C} that is not positive definite (two negative
eigenvalues). We do not see a simple way to avoid this flaw;
however, it should be inconsequential in practice as 
only the density matrices of \textbf{A} and \textbf{B} are
used to compute the ISAPT corrections and, for the 
purpose of recomputing the occupied orbital spaces
${\cal B}_{\mathbf A}$ and ${\cal B}_{\mathbf B}$, the
repartitioning only changes the embedding potential at the
more distant interfragment boundary.

\section{Results and discussion}\label{sec:discussion}

Our improved ISAPT algorithms have been implemented in a development version of the Psi4 software package \cite{Smith:20}.
The modified Psi4 code is available on GitHub at\\ \verb+https://github.com/konpat/psi4/tree/isapt+.
The interaction energy components within a number of representative molecules are analyzed and compared to the previously proposed ISAPT method by Parrish et al. \cite{Parrish:15} In particular, pentanediol isomers, n-heptane, 2,4-dimethylpentane, and folded and unfolded bicyclic N-arylimide based molecular balances (halogens: Cl, Br, and I interact noncovalently with aromatic fragments: benzene, phenanthrene, pyrene, and ethylene)\cite{Sun:17} are chosen for the study. All geometries are optimized at the MP2 level of theory employing the aug-cc-pVDZ basis set. To examine the accuracy of energy contributions, single-point ISAPT energy calculations are performed using three orbital bases, aug-cc-pVDZ, aug-cc-pVTZ, and aug-cc-pVQZ (aDZ, aTZ, aQZ for short). For bulky molecules, the N-arylimide balances, the systems are optimized using the B3LYP-D3(BJ) method \cite{Grimme:11} with the aDZ basis set (with cc-pVDZ-pp on the heavy iodine atom). The resulting ISAPT interaction energies are computed using the aDZ basis set, with cc-pVTZ centered on the iodine atom. Overall, each molecular system is tested with seven link assignment options: C (original), SAO0, SAO1, SAO2, SIAO0, SIAO1, and SIAO2, with and without orthogonalizing the link orbitals to the fragment occupied space.

\begin{figure}
\begin{minipage}{.30\textwidth}
\centering
\includegraphics[angle=-15, width=.9\textwidth,height=2.75cm]{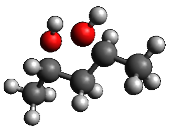}
\caption*{P242}
\end{minipage}
\begin{minipage}{.30\textwidth}
\centering
\includegraphics[angle=14, width=1\textwidth]{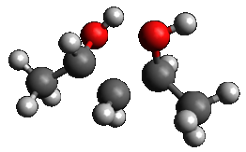}
\caption*{P244}
\end{minipage}
\begin{minipage}{.30\textwidth}
\centering
\includegraphics[angle=20, width=.9\textwidth]{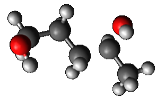}
\caption*{P142}
\end{minipage}
\begin{minipage}{.30\textwidth}
\centering
\includegraphics[width=.96\textwidth]{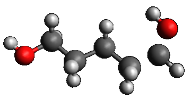}
\caption*{P156}
\end{minipage}
\begin{minipage}{.30\textwidth}
\centering
\includegraphics[angle=-8, width=.9\textwidth]{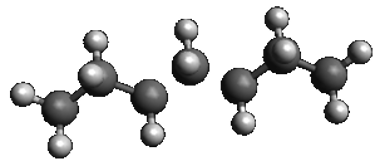}
\caption*{C73}
\end{minipage}
\begin{minipage}{.30\textwidth}
\centering
\includegraphics[angle=-3, width=.68\textwidth]{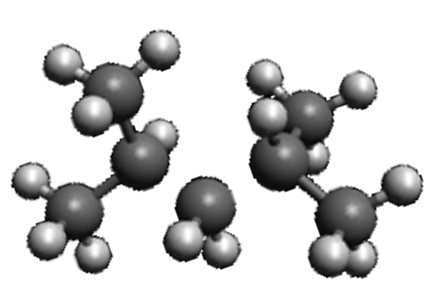}
\caption*{C7B24}
\end{minipage}
\caption{Some illustrative fragmentation patterns of  2,4-, 1,4-, and 1,5-pentanediol, n-heptane, and 2,4-dimethylpentane considered in this work. A missing bond signifies a fragment boundary.}
\label{fig:pentanediols_n_heptane}
\end{figure}

\subsection{Pentanediol isomers}

The 2,4-, 1,4-, and 1,5-pentanediol molecules, featuring the OH$\cdots$O distances ranging from 2.40 \text{\normalfont\AA} (an intramolecular hydrogen bond) to 6.30 \text{\normalfont\AA}, serve as convenient illustrative examples for a range of polar noncovalent intramolecular interactions.  
Four, six, and nine fragmentation patterns of the 2,4-, 1,4-, and 1,5-pentanediol systems, respectively, were studied.
A few examples (P242, P244, P142, and P156) of these fragmentation patterns are presented in Fig. \ref{fig:pentanediols_n_heptane}; the remaining ones are shown in the Supporting Information.

\begin{figure}[!htb]
\begin{minipage}{1\textwidth}
\centering
\includegraphics[width=1\textwidth]{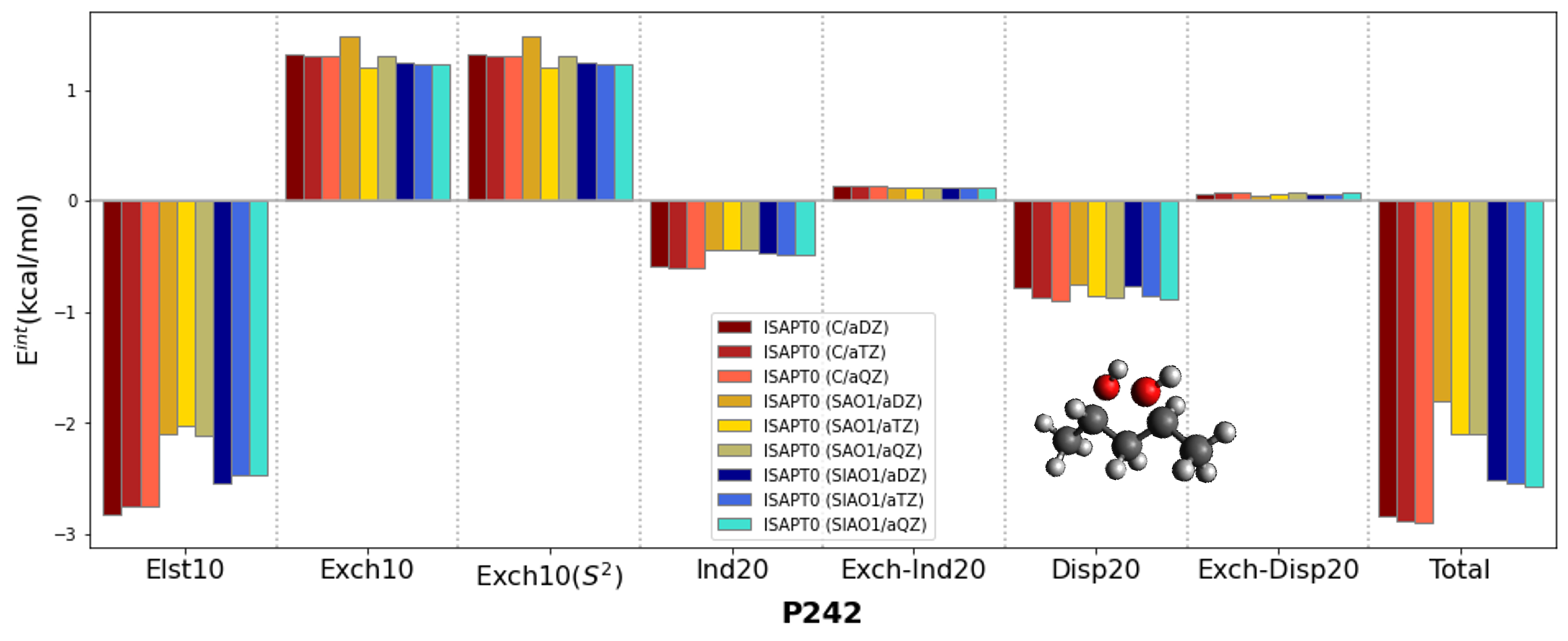}
\caption{ISAPT energy components in 2,4-pentanediol (P242 fragmentation pattern) computed using the aXZ bases (X=D, T, Q) and three link assignments: C, SAO1, and SIAO1.}
\label{fig:P242-plot}
\end{minipage}
\end{figure}

The P242 model, well handled by the original ISAPT(C) approach, assists in the verification of the implemented methods by comparing to the data in Ref.\citenum{Parrish:15}. 
According to Fig. \ref{fig:P242-plot}, the ISAPT energy components computed by the SIAO1 method have the lowest absolute differences relative to the original C variant.  The electrostatic term has a MAD, mean absolute difference, of 0.28 kcal/mol with bases aDZ, aTZ, and aQZ. This MAD amounts to 0.70 kcal/mol for the SAO1 method.  Other energy contributions are also in good agreement between different fragmentation patterns:  the SAO1 and SIAO1 induction energies deviate from the C variant by 0.12--0.17 kcal/mol, the first-order exchange term varies by about 0.07 kcal/mol, and differences in the dispersion term do not exceed 0.02 kcal/mol, much smaller than the differences between basis sets.  

\begin{figure}[!htb]
\begin{minipage}{\textwidth}
\centering
\includegraphics[width=\textwidth]{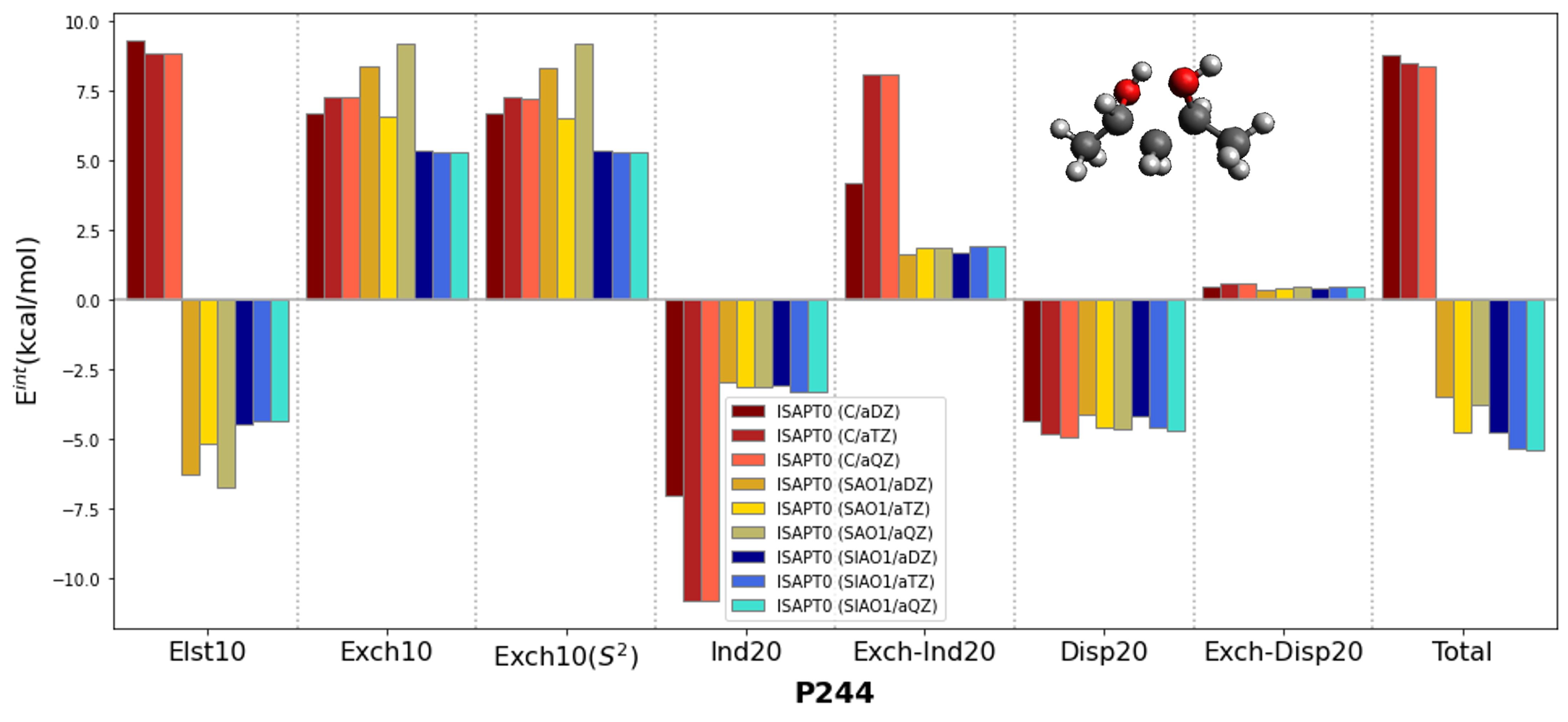}
\end{minipage}
\caption{ISAPT energy components in 2,4-pentanediol (P244 fragmentation pattern) computed using the aXZ bases (X=D, T, Q) and three link assignments: C, SAO1, and SIAO1.}
\label{fig:P244-plot}
\end{figure}

\begin{figure}[!htb]
\begin{minipage}{\textwidth}
\centering
\includegraphics[width=\textwidth]{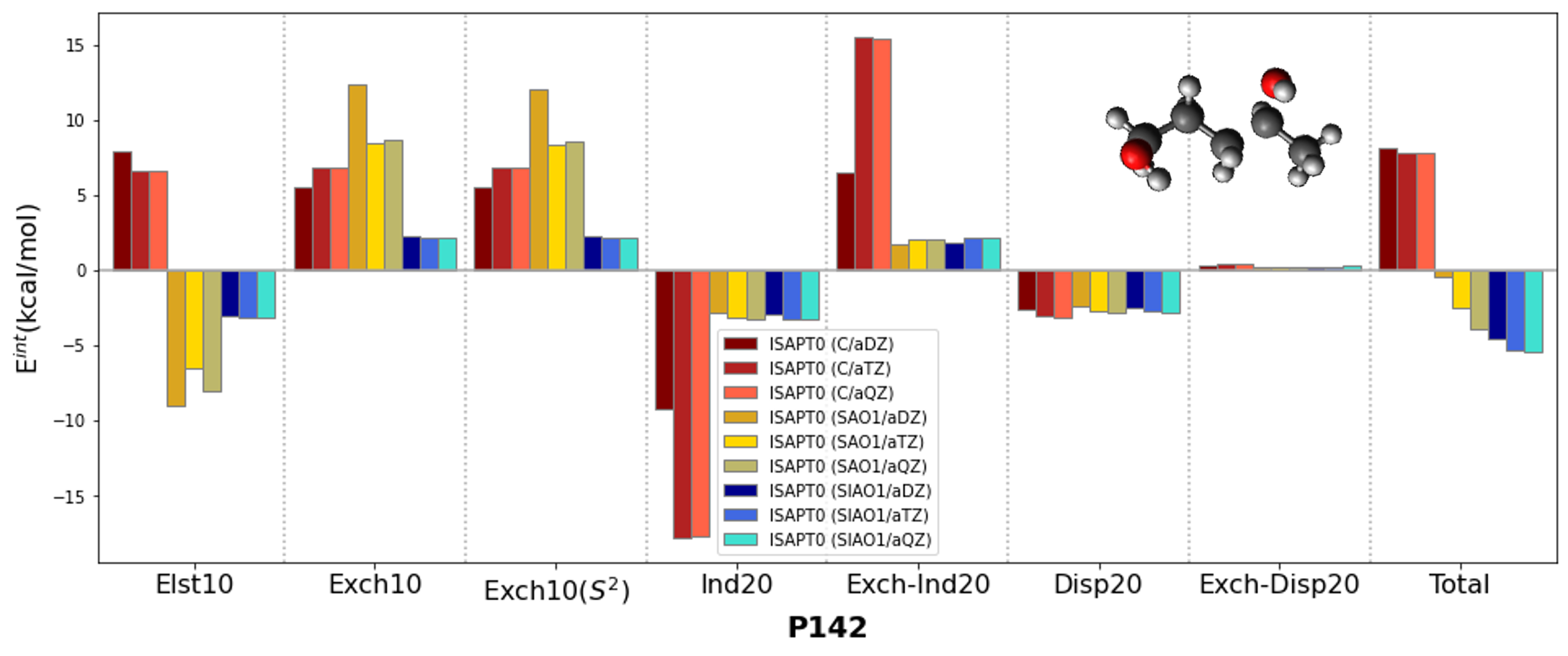}
\end{minipage}
\caption{ISAPT energy components in 1,4-pentanediol (P142 fragmentation pattern) computed using the aXZ bases (X=D, T, Q) and three link assignments: C, SAO1, and SIAO1.}
\label{fig:P142-plot}
\end{figure}

\begin{figure}[!htb]
\begin{minipage}{\textwidth}
\centering
\includegraphics[width=\textwidth]{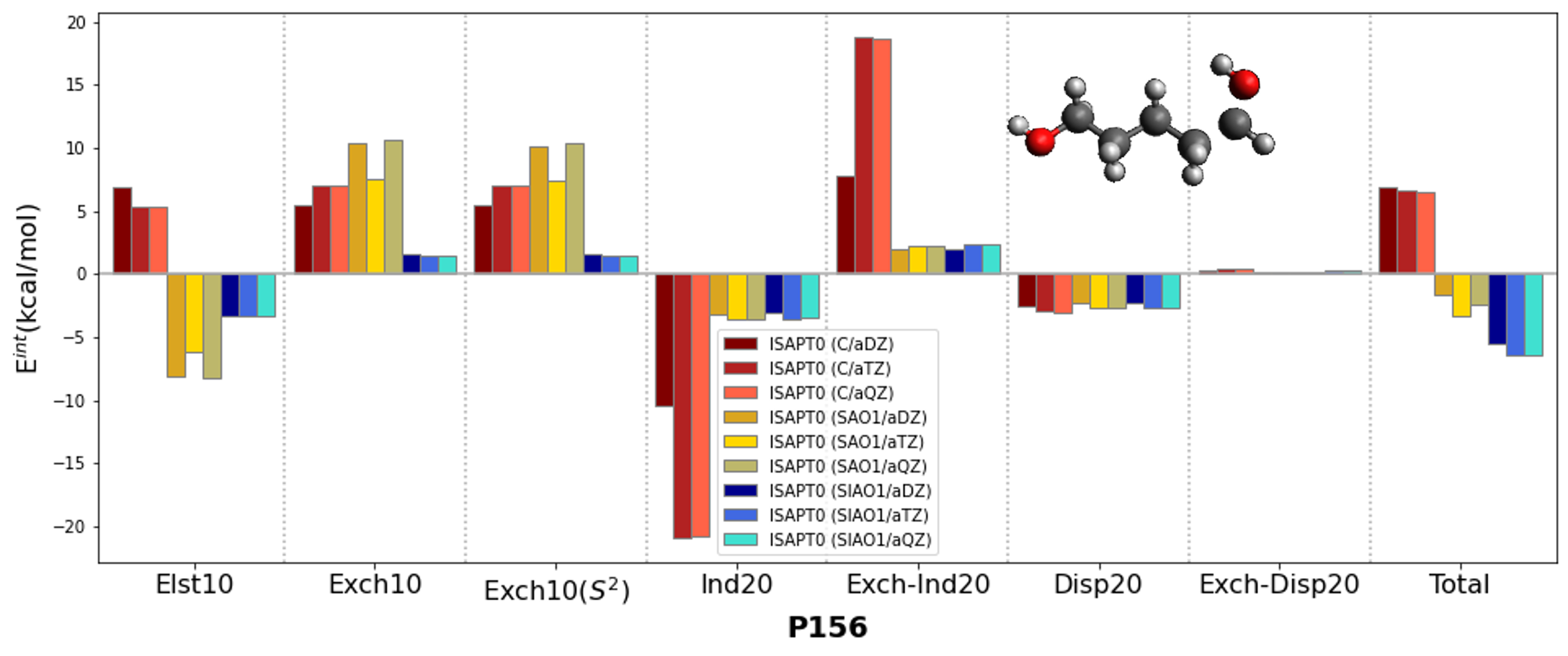}
\end{minipage}
\caption{ISAPT energy components in 1,5-pentanediol (P156 fragmentation pattern) computed using the aXZ bases (X=D, T, Q) and three link assignments: C, SAO1, and SIAO1.}
\label{fig:P156-plot}
\end{figure}

On the other hand, for most of the other fragmentation patterns, including the minimal linker arrangement of 2,4-pentanediol (P244) as well as related short-linker configurations for other isomers (P142 and P156), the electrostatic energy in the original ISAPT(C) variant is strongly repulsive.
This does not make physical sense, as one expects either strong electrostatic attraction due to favorable dipole-dipole interaction (P244) or weak attraction due to charge penetration (P142 and P156).
As stated before, we hypothesize that the observed problematic behavior of ISAPT(C) electrostatics is connected to the unphysical dipole moments emerging at the \textbf{A}-\textbf{C} and \textbf{B}-\textbf{C} fragment boundaries.
The results with the new ISAPT link assignments (Figs. \ref{fig:P244-plot}-\ref{fig:P156-plot})
indicate that the SAO1 and SIAO1 electrostatic energy has shifted to a physically justified negative value.  
Furthermore, while the second-order induction and exchange-induction components in original ISAPT(C) are both very large in magnitude and cancel each other to a large extent, the respective SAO1 and SIAO1 terms are much smaller.
The large cancellation between induction and exchange-induction effects is well known from conventional intermolecular SAPT (and is related to the overall divergence of the perturbation series \cite{Patkowski:04,Misquitta:13}):
the unsymmetrized (polarization) expansion allows the two molecules to overpolarize each other, an effect that has to be cancelled by the exchange terms brought about by enforcing the full permutational symmetry of the wavefunction.
The observation that, in ISAPT, such an overpolarization is made particularly egregious by the original C link assignment, once again suggests the unphysical dipole moments at interfragment boundaries, leading to an artificially enhanced electric field, as the culprit. 
As far as other ISAPT energy terms are concerned, the first-order exchange is relatively consistent between the C and SAO1 schemes but reduced in the SIAO1 one while the dispersion and exchange-dispersion energies are remarkably consistent across all three link assignments.

Figures \ref{fig:P244-plot}-\ref{fig:P156-plot} show that the link reassignment proposed in this work significantly improves the ISAPT energy components that are problematic in original ISAPT(C).
We postulate that this improvement stems from a substantial reduction of the artificial dipoles arising from the nonsymmetric carbon atoms next to the \textbf{A}-\textbf{C} and \textbf{B}-\textbf{C} linking bonds. 
Indeed, Table \ref{table:dipole} shows that the SAO1 and SIAO1 dipole moments on fragments \textbf{A} and \textbf{B}, for different pentanediol isomers and fragmentation patterns, are about 15--20$\%$ smaller in magnitude than the fragment dipole moments from the original ISAPT0, and are consistent with respect to the basis set. As both fragments are obviously polar, we do not expect a reduction of dipole moments to near-zero in pentanediol molecules. 
However, it appears that the redistribution of link bonds to the corresponding IHOs on each fragment has crucially reduced the imbalance of the electron densities on the linking atoms, leading to more meaningful interaction energies.

\begin{table}[!h]
\caption{Magnitude of the HF dipole moments (in a.u.) for different molecules and their fragments in the aDZ, aTZ, and aQZ bases.}
\label{table:dipole}
\centering
\nprounddigits{3}
\npdecimalsign{.}
\begin{tabular}{c|c|n{2}{3}|n{2}{3}n{3}{3}n{3}{3}|n{2}{3}n{3}{3}n{3}{3}}
\hline
\multirow{2}{*}{\textbf{System}} & \multirow{2}{*}{\textbf{Basis}} &\multicolumn{1}{c|}{\multirow{2}{*}{\textbf{ABC}}} & \multicolumn{3}{c|}{Fragment \textbf{A}} &\multicolumn{3}{c}{Fragment \textbf{B}}  \\\cline{4-9}
&&& \multicolumn{1}{c}{C} & \multicolumn{1}{c}{SAO1} & \multicolumn{1}{c|}{SIAO1}& \multicolumn{1}{c}{C} & \multicolumn{1}{c}{SAO1} & \multicolumn{1}{c}{SIAO1}\\ 
\hline\hline
\multirow{3}{*}{P242} & {aDZ} & 1.05661572 & 0.95106745 & 0.6698701 & 0.71579365 & 0.93604285 & 0.66604209 & 0.69080439 \\\cline{2-9}
& {aTZ} & 1.05620391 & 0.95018643 & 0.64416662 & 0.71363284 & 0.93519195 & 0.63309439 & 0.68874972\\\cline{2-9}
& {aQZ} & 1.05597091 & 0.95028123 & 0.65700271 & 0.71349785 & 0.93526654 & 0.63007153 &   0.68863294\\
\hline
\multirow{3}{*}{P244} & {aDZ} & 1.05661572 & 0.83253202 & 0.69900568 & 0.68826343 & 1.31162816 & 0.72019653 & 0.9219332 \\\cline{2-9}
&{aTZ} & 1.05620391 & 0.83290891 & 0.70423936 & 0.68686496 & 1.31261494 & 0.76490003 & 0.92082409 \\\cline{2-9}
& {aQZ} & 1.05597091 & 0.83295935 & 0.74357754 & 0.68680132 & 1.31253506 & 0.73982609 & 0.92067303 \\
\hline
\multirow{3}{*}{P142} & {aDZ} & 1.04032399 & 1.31867929 & 0.67577229 & 0.91475776 & 0.89284163 & 0.62450235 & 0.69671412 \\\cline{2-9}
& {aTZ} & 1.03954536 & 1.31947392 & 0.68042191 & 0.91301695 & 0.89196181 & 0.61065677 & 0.69438606  \\\cline{2-9}
& {aQZ} & 1.03918211 & 1.3193482 & 0.6383017 & 0.91271335 & 0.89201216 & 0.62824062 & 0.69419065\\
\hline
\multirow{3}{*}{P156} & {aDZ} & 0.95855327 & 1.18613748 & 0.630206 & 0.89327451 & 0.89303977 & 0.62541863 & 0.7046575\\\cline{2-9}
&{aTZ} & 0.95760761 & 1.18483444 & 0.66651976 & 0.88946703 & 0.8922242 & 0.61724158 & 0.70188454\\\cline{2-9}
&{aQZ} & 0.95748611 & 1.18442627 & 0.60362316 & 0.88897949 & 0.89211448 & 0.61160816 & 0.70149434\\
\hline
\multirow{3}{*}{C73} & {aDZ} &   0.04677840 & 0.80822728 &  0.09544139 & 0.38884191 & 0.81335938 & 0.09714934 & 0.39154473 \\\cline{2-9}
& {aTZ} & 0.04597183 & 0.80960883 & 0.04806767 & 0.38765703 & 0.81485504 & 0.05574455 & 0.39054173 \\\cline{2-9}
& {aQZ} & 0.04597183  & 0.80966069 & 0.10961002 & 0.38761189 & 0.81492064 & 0.12627977 &  0.39051227\\
\hline
\multirow{3}{*}{C75} & {aDZ} & 0.04677840 & 0.78829343 &  0.11541527 &  0.33931031 & 0.81335938 &  0.08035142 & 0.35172353\\\cline{2-9} 
&{aTZ} & 0.04597183 & 0.79019722 & 0.04729118 &  0.33911540 & 0.81485504 & 0.03667895 & 0.35107107 \\\cline{2-9} 
&{aQZ} & 0.04597183 & 0.79026574 & 0.04116558 & 0.33905586 & 0.81492063 & 0.11731597 &  0.35100669\\
\hline
\multirow{3}{*}{C7B24} & {aDZ} & 0.04496584 & 0.79881496 & 0.09110603 & 0.38529185 & 0.79876027 & 0.09092715 & 0.3852933\\\cline{2-9} 
&{aTZ} & 0.04407461 & 0.80023499 & 0.17754943 & 0.3844555 & 0.8001835 & 0.17745038 & 0.38445962\\\cline{2-9}
& {aQZ} & 0.04403493 & 0.80025348 & 0.10014621 & 0.3843893 & 0.80020164 & 0.10004333 & 0.38439299\\
\hline
\end{tabular}
\end{table}

\begin{figure}[!htb]
\begin{minipage}{\textwidth}
\centering
\includegraphics[width=.95\textwidth]{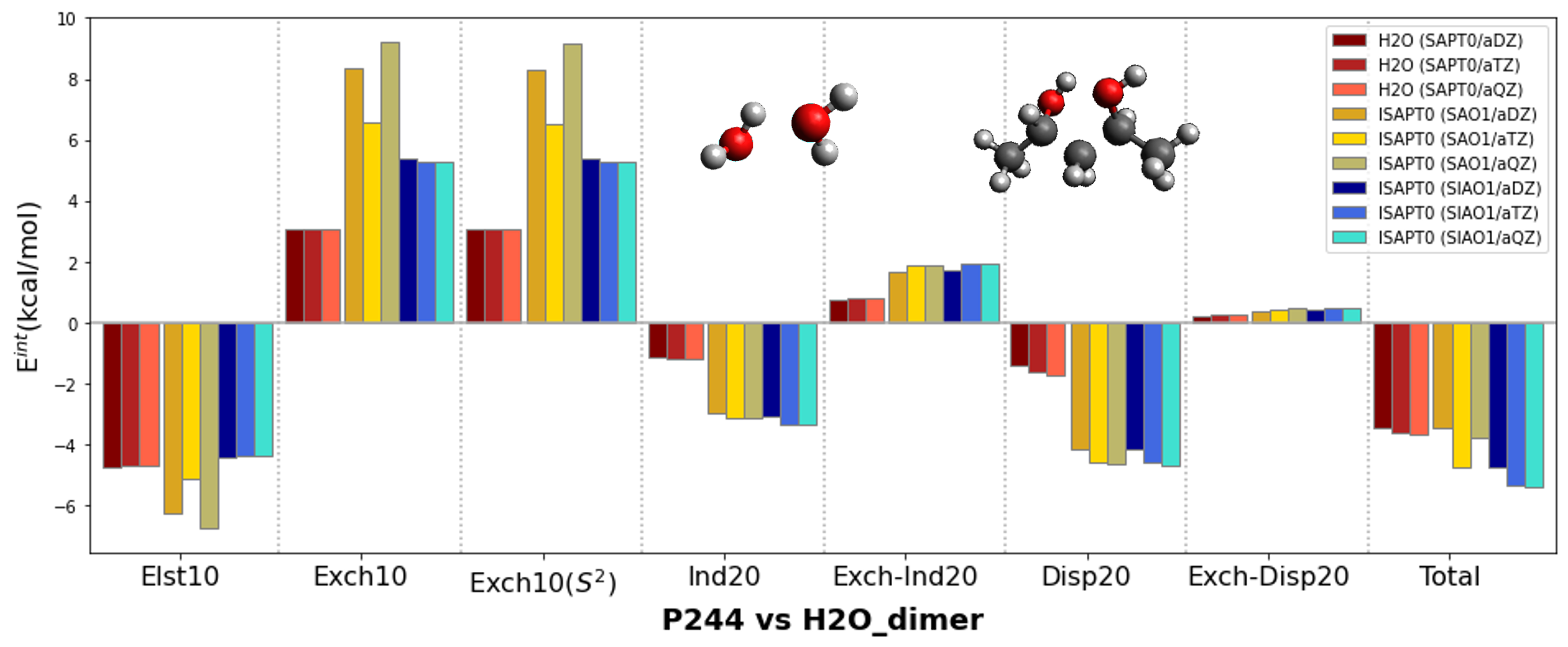}
\caption{Comparison of energy components between the P244 fragmentation pattern of 2,4-pentanediol (computed with ISAPT/SAO1 and ISAPT/SIAO1) and the corresponding water dimer structure exhibiting the same hydrogen-bonded arrangement (computed with standard SAPT0).}
\label{fig:P244-h2odimer}
\end{minipage}
\end{figure}

Overall, with the new SAO1 and SIAO1 variants, the dominant attractive contributions to the intramolecular interaction energy in 2,4-pentanediol are electrostatics and dispersion as expected for a hydrogen bonded system.
The overall interaction energy for the P244 fragmentation pattern, Fig. \ref{fig:P244-plot}, is attractive as expected, amounting to $-3.48$ kcal/mol for SAO1/aDZ and $-4.75$ kcal/mol for SIAO1/aDZ, improving from the repulsive 8.80 kcal/mol value obtained with original ISAPT0(C)/aDZ. 
The SAO1 and SIAO1 variants predict that the fragments in P244 exhibit significantly stronger bonding than the fragments in P242 (Fig. \ref{fig:P242-plot}), which is consistent with the addition of aliphatic chains that amplify the attractive dispersion and charge penetration effects.
Both models can also be compared with an analogous hydrogen bonded configuration of the water dimer,
involving the same OH$\cdots$OH geometry with the remaining hydrogens added in the direction of the O-C bonds in 2,4-pentanediol at MP2/aDZ-optimized distances (Fig.~\ref{fig:P244-h2odimer}).
One can see that the attractive electrostatic energy obtained from the intramolecular SIAO1 method is in good agreement with the water dimer value computed with standard intermolecular SAPT0.
This agreement is accidentally too good: both electrostatic terms primarily stem from the dipole-dipole interaction of the very similar polar fragments, but 2,4-pentanediol includes more charge penetration due to the hydrocarbon chains.
Indeed, the presence of the additional aliphatic chains in P244 increases both the interfragment density overlap (thus leading to larger first-order exchange) and the polarizability (enhancing the induction and dispersion terms).

When the basis set is enlarged from aDZ to aQZ, the ISAPT(SIAO1) energy components are highly consistent. As expected, the dispersion and exchange-dispersion corrections increase in magnitude as the basis set increases while other, uncorrelated corrections change very little. 
This observation is in perfect agreement with standard intermolecular SAPT, where the slow basis set convergence of dispersion and exchange-dispersion corrections is well documented and the possible remedies include midbond functions \cite{Williams:95} and the explicitly correlated F12 approach \cite{Kodrycka:21}. 
Somewhat disappointingly, the same stable convergence pattern does not apply to the SAO1 link assignment.
Conversely, Figs.~\ref{fig:P242-plot}-\ref{fig:P156-plot} indicate that the basis set convergence of the ISAPT(SAO1) electrostatic and first-order exchange terms is slow and erratic, and these convergence issues carry on to the total interaction energies.
This ISAPT(SAO1) basis set instability is the main reason why we recommend the SIAO1 variant, which never suffers from such convergence issues, for all practical applications.

\begin{figure}[!htb]
\begin{minipage}{\textwidth}
{\bf (a)} \hfil {\bf (b)} \hfil \\
\includegraphics[width=0.48\textwidth]{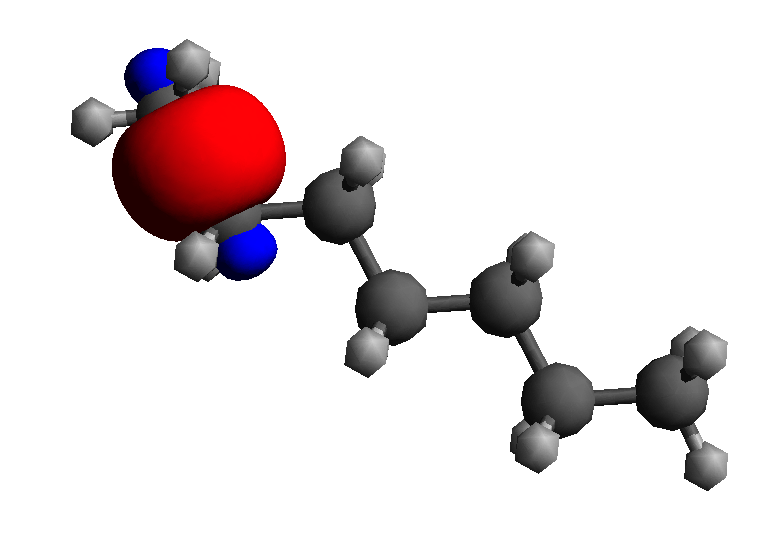}\hfil\includegraphics[width=0.48\textwidth]{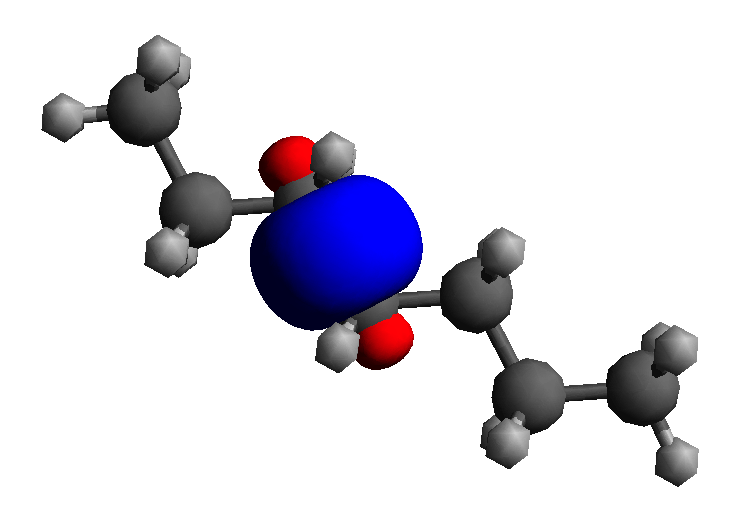}\\
{\bf (c)} \hfil {\bf (d)} \hfil \\
\includegraphics[width=0.48\textwidth]{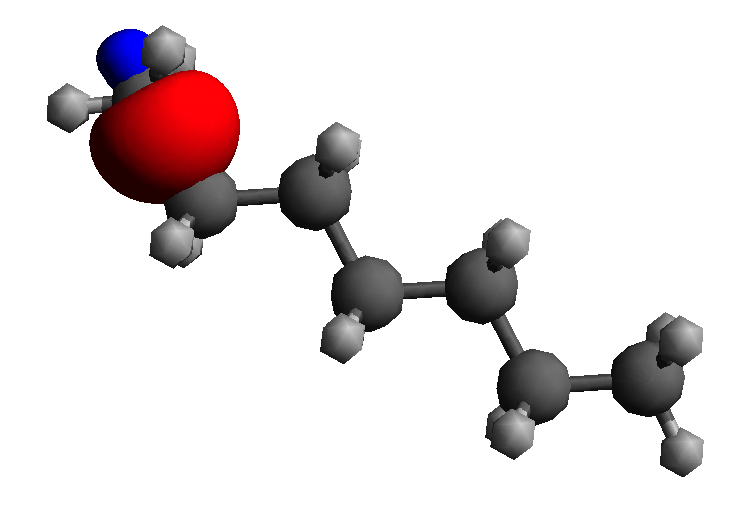}\hfil\includegraphics[width=0.48\textwidth]{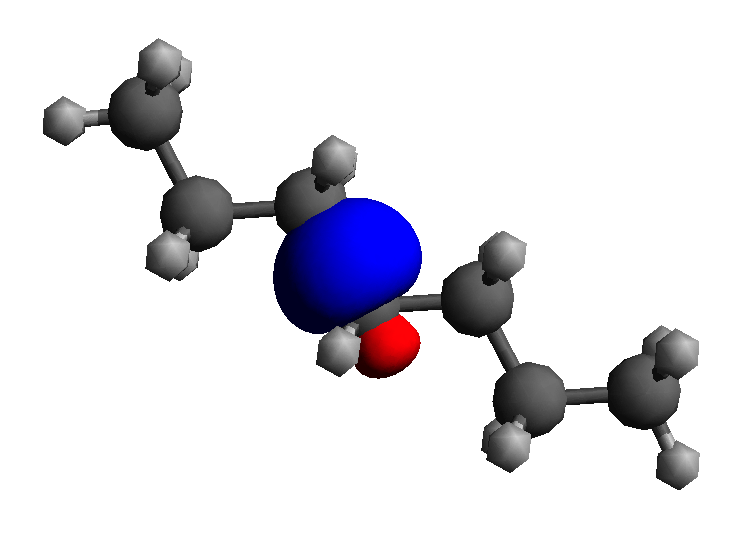}\\
{\bf (e)} \hfil {\bf (f)} \hfil \\
\includegraphics[width=0.48\textwidth]{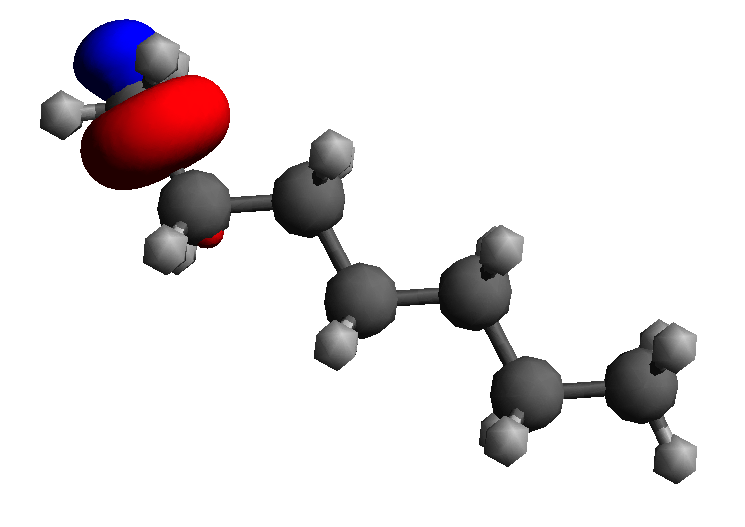}\hfil\includegraphics[width=0.48\textwidth]{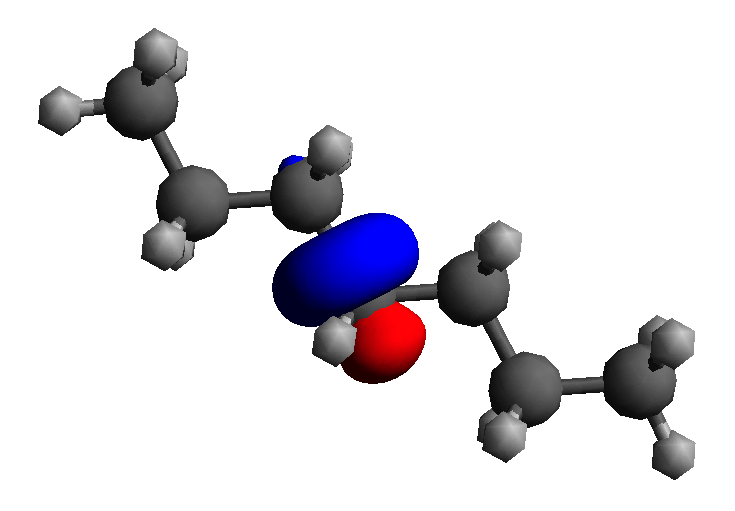}\\
\end{minipage}
\caption{Orbital contours (at the same isosurface value of 0.05) of the complete link IBOs from ISAPT(C) (panels {\bf (a)} and {\bf (b)}) and the reassigned link IHOs from ISAPT(SAO1) (panels {\bf (c)} and {\bf (d)}) and from ISAPT(SIAO1) (panels {\bf (e)} and {\bf (f)}). The system is n-heptane in the C74 fragmentation pattern, and the basis set is aDZ.}
\label{fig:orbitals_heptane}
\end{figure}

Some insights into the different basis set behavior of ISAPT(SAO1) and ISAPT(SIAO1) can be obtained by examining the corresponding link orbitals (IHOs). 
An example comparison of the link IBOs with the IHOs resulting from both schemes, for the n-heptane molecule and its fragmentation pattern later referred to as C74, is presented in Fig.~\ref{fig:orbitals_heptane}.
For this example, we note that both schemes lead to IHOs that roughly resemble the carbon sp$^3$ hybrid orbitals pointing towards the other fragment, however, the SIAO1 orbitals appear to be more localized to the respective \textbf{A}/\textbf{B} fragments (note that all panels in Fig.~\ref{fig:orbitals_heptane} use the same isosurface value).
This would suggest that the two IHOs $\chi_x'$ and $\chi_y'$ should have smaller overlap for SIAO1 than for SAO1 (as before, the primes signify that one iteration towards self-consistency of orbital spaces and link IHOs has been performed within the SAO1/SIAO1 schemes).
Indeed, we observe that the $\langle\chi_x'|\chi_y'\rangle$ overlap integral tends to be much smaller in magnitude for SIAO1 than for SAO1. 
For example, in the 2,4-pentanediol system, the P244 fragmentation scheme, and the aDZ, aTZ, and aQZ basis sets, $|\langle\chi_x'|\chi_y'\rangle|$ amounts to 0.133, 0.0632, and 0.0724, respectively, for SAO1 and 3.02x10$^{-4}$, 3.09x10$^{-4}$, and 3.11x10$^{-4}$, respectively, for SIAO1.
Formally, neither SIAO1 nor SAO1 link hybrids are strictly localized on the \textbf{A}/\textbf{B} fragments: the IAOs have small tails on other atoms, and the orthogonalization does not strictly preserve the localization. 
The reason why the SIAO1 link hybrids are more strongly localized than the SAO1 ones is that the minimal basis used to construct IAOs \cite{Knizia:13} does not give them the flexibility to improve the basis set description on atoms other than their center.
On the contrary, in a sufficiently large AO basis, the functions centered on one atom actively improve the basis set description around other atoms, although this improvement varies erratically from basis to basis.
As a result, the SAO projection scheme leaves out larger tails than the SIAO one, but these tails are not stable with the basis set.
In our opinion, this behavior leads to both a larger $|\langle\chi_x'|\chi_y'\rangle|$ overlap and a larger basis set instability for the SAO1 link assignment scheme relative to the SIAO1 one.

As the OH$\cdots$O interacting distance increases when going from 2,4- to 1,4- and 1,5-pentanediols, the electrostatic contribution in a primarily dipole-dipole interaction decreases. Among the 1,4-pentanediol models, the P142 partitioning reports the most favorable electrostatic energy (Fig. \ref{fig:P142-plot}), which amounts to $-3.04$ kcal/mol for SIAO1 and $-9.01$ kcal/mol for SAO1 using the aDZ basis.  A notable reduction in the induction term compared to the original ISAPT(C) variant occurs, leading to a net interaction energy of $-4.50$ kcal/mol and $-0.38$ kcal/mol in SIAO1 and SAO1,  respectively.  Interestingly,  a similar trend is observed in the P156 partitioning of the 1,5-pentanediol system (Fig. \ref{fig:P156-plot}), yet the electrostatic term can more likely be attributed to charge penetration in this case. 
The 1,5-pentanediol series shows that the fragments are mainly bound by the electrostatic force, along with a roughly equal contribution from the induction and dispersion terms that sum to the net attractive energy. 

\subsection{Alkanes}

Due to the nonpolarity of the C-C bond and the negligible electronegativity difference involved in the C-H bonding, intramolecular interactions in hydrocarbons are dominated by London dispersion forces.
Alkanes are excellent models for analyzing nonbonded intramolecular interactions, as it is those interactions that result in a higher thermodynamic stability of branched alkanes relative to their linear isomers \cite{Steinmann:10a}.
Thus, linear and branched seven-carbon alkanes, n-heptane and 2,4-dimethylpentane (structures C73 and C7B24 in Fig. \ref{fig:pentanediols_n_heptane}, respectively) are chosen as models to study the performance of different ISAPT variants. 
It should be stressed that a single fragmentation of these alkanes is not enough to quantify the entire nonbonded intramolecular interaction, as some of the relevant interacting groups inevitably end up on the same fragment.
However, the fragmentation patterns in C73 and C7B24 are designed to alleviate this issue as, in both cases, fragments \textbf{A} and \textbf{B} are propyl groups (1-propyl or 2-propyl). 
Thus, the \textbf{A}-\textbf{B} interaction misses the important 1,3-methyl-methyl stabilizing effects (protobranching \cite{Wodrich:07}) within each propyl group, but it misses the same number of such effects in both cases.

Both n-heptane and 2,4-dimethylpentane are nonpolar molecules, and their fragments obtained by cutting through C-C bonds should be nonpolar as well.
However, as shown in Table \ref{table:dipole}, when the entire link IBO is assigned to fragment \textbf{C}, the noncovalently interacting fragments \textbf{A} and \textbf{B} acquire large dipole moments.
As stated above, the origin of those unphysical dipole moments is the unbalanced charge distribution around the linking carbon atoms, where the electrons occupying only three out of the four $sp^3$ orbitals belong to the fragment \textbf{A}(\textbf{B}).
Fortunately, according to Table \ref{table:dipole}, the magnitude of the fragment dipole moments is strongly reduced in the new approaches, by 85--90$\%$ for SAO1 and 50--55$\%$ with SIAO1. 
In the SIAO1 case, this reduction is perfectly consistent in all tested basis sets, while the SAO1 dipole moments show some basis set fluctuations in line with the energy components. 
Thus, while the link bond reassignment proposed here does not completely eliminate the spurious multipoles at the interfragment boundary, it reduces their magnitude significantly. 

\begin{figure}[!htb]
\begin{minipage}{\textwidth}
\centering
\includegraphics[width=\textwidth]{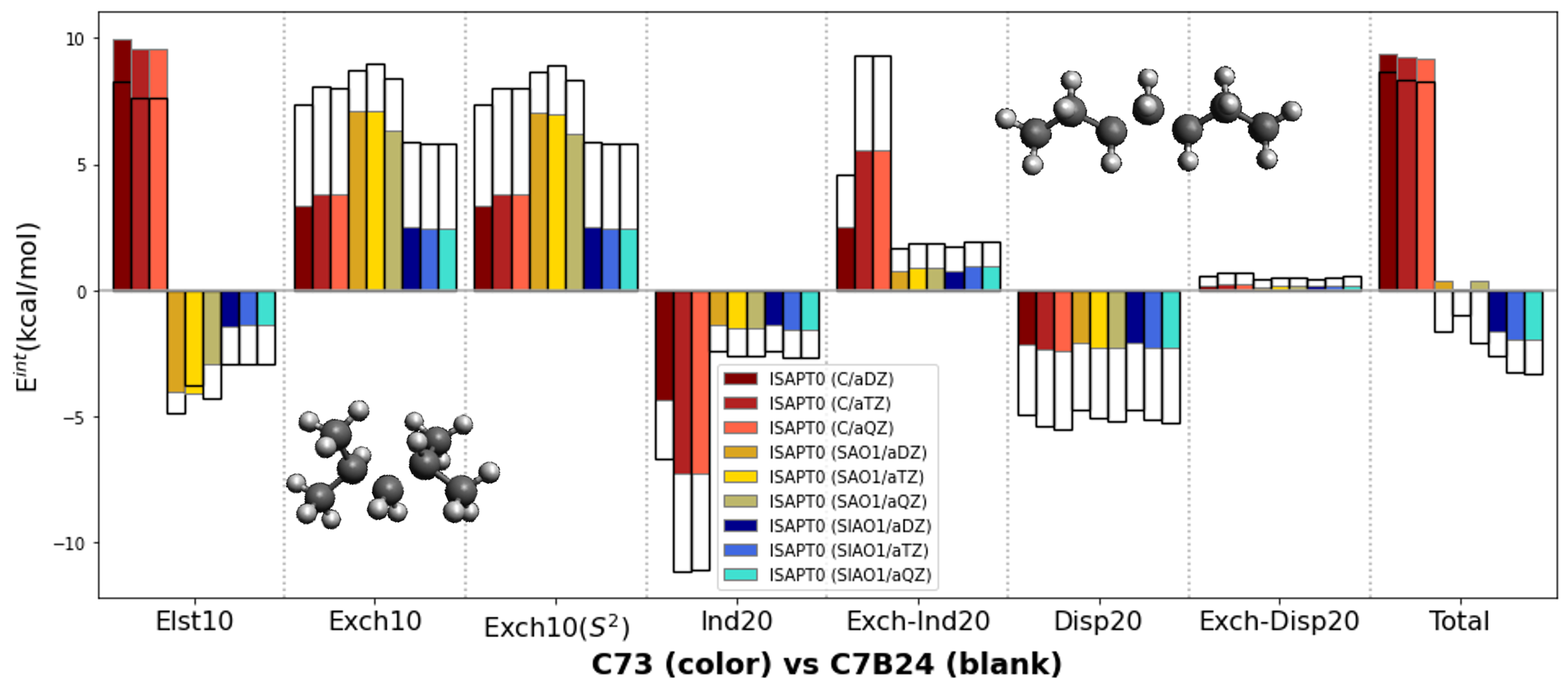}
\end{minipage}
\caption{Comparison of the ISAPT0 energy components between n-heptane (C73, colored bars) and its branched isomer, 2,4-dimethylpentane (C7B24, empty bars) computed with the C, SAO1, and SIAO1 variants.}
\label{fig:C73-C7B24-plot}
\end{figure}

We now turn to examining the ISAPT energy contributions for n-heptane and 2,4-dimethylpentane as a function of the link assignment, focusing on the difference between two systems which may shed light on the physical origins of the branched alkane stabilization.
A glimpse at both sets of data (Fig. \ref{fig:C73-C7B24-plot}) once again shows that the default ISAPT(C) variant does not provide physically meaningful results: the electrostatic energy is strongly repulsive while the induction and exchange-induction terms are very large in magnitude and strongly vary between aDZ and aTZ.
The SAO1 link reassignment once again alleviates the issue of repulsive electrostatics.
However, besides a fairly slow (but not terrible) basis set convergence, one sees another troubling property that puts the usefulness of ISAPT(SAO1) into doubt: the total C73 interaction energy is slightly repulsive in all tested basis sets. 
The SIAO1 link reassignment eliminates both of those issues, providing meaningful and quickly convergent ISAPT energy contributions and predicting a stronger interfragment stabilization for the branched isomer relative to the linear one.
Thus, only the ISAPT(SIAO1) variant is suitable for providing meaningful insights into the nonbonded interactions enhancing the thermodynamic stability of 2,4-dimethylpentane relative to n-heptane.

The ISAPT(SIAO1) data in Fig. \ref{fig:C73-C7B24-plot} indicate that the overall attractive interaction energy is predominantly controlled by the dispersion contribution and secondarily by the electrostatic term, but the latter effect is more than counterbalanced by the repulsive first-order exchange.
The 2,4-dimethylpentane system exhibits roughly twice the intramolecular electrostatic and dispersion energy of the n-heptane model ($-2.95$ and $-4.76$ kcal/mol versus $-1.43$ and $-2.09$ kcal/mol, respectively, in the aDZ basis set).
This illustrates why branched alkanes are more thermodynamically stable than linear alkanes with the same carbon content \cite{Steinmann:10a} --- the contribution to the fragments’ electron densities from multiple closely spaced methyl groups enhances the charge penetration term in the electrostatic energy as well as leads to a larger dispersion energy. 
Obviously, the larger interfragment overlap in the branched system increases the first-order exchange repulsion as well, and the overall change in interaction energy is the net result of the additional stabilizing terms in electrostatics and dispersion and destabilizing exchange contributions. 

\begin{figure}[!htb]
\begin{minipage}{\textwidth}
\centering
\includegraphics[width=\textwidth]{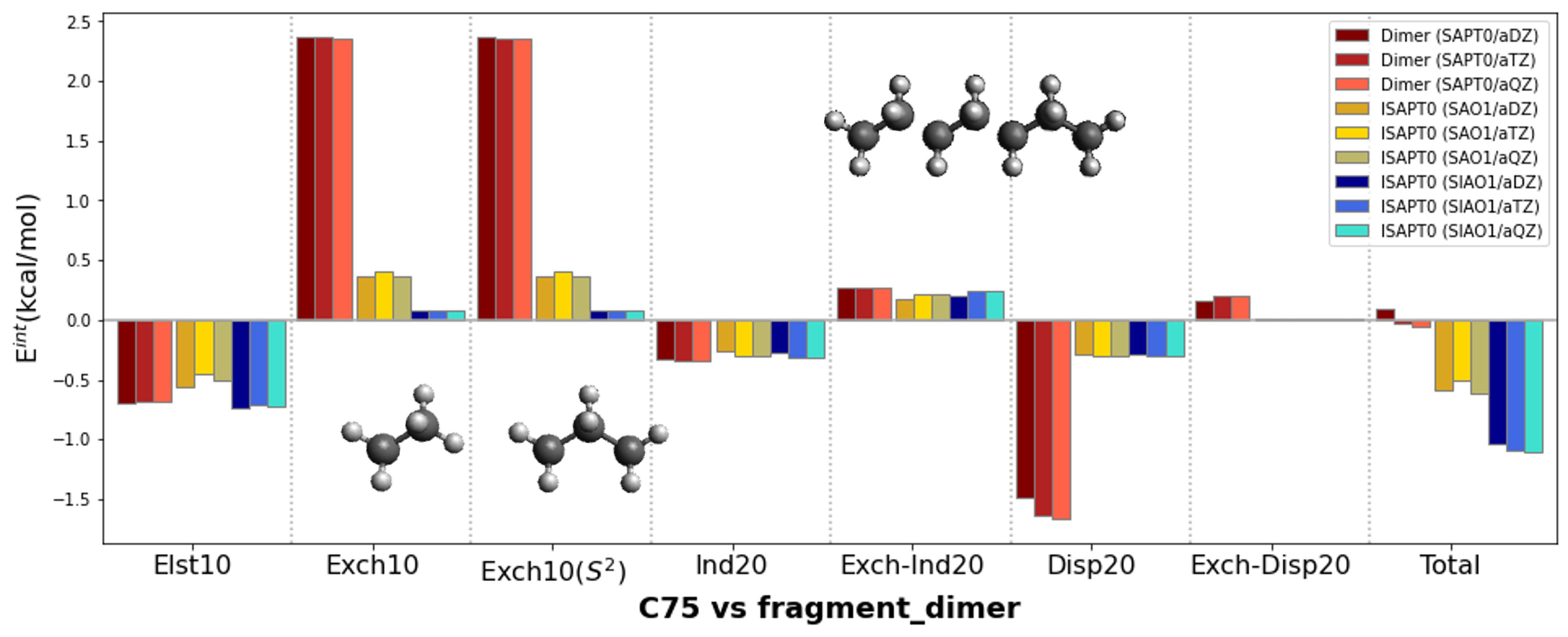}
\caption{Comparison of energy components between the C75 model of n-heptane (computed with the ISAPT0/SAO1 and ISAPT0/SIAO1 variants) and the related intermolecular propane-ethane dimer (computed with standard SAPT0) in the aXZ bases, X=D, T, Q. The locations of two additional hydrogens in the dimer calculation are explicitly optimized at the MP2/aDZ level.}
\label{fig:C75_frag-dimer}
\end{minipage}
\end{figure}

In addition, the intramolecular energy contributions in n-heptane can be compared to the standard SAPT terms for a related intermolecular interaction involving fragments \textbf{A} and \textbf{B} capped with hydrogen atoms. 
The C73 model with the small -CH$_2$- linking fragment is not suitable for this purpose as the capping hydrogens would end up too close to each other.
Therefore, we switch to the C75 model with a slightly bigger -CH$_2$CH$_2$- linking fragment.
For this model, the interacting fragments \textbf{A} and \textbf{B}  with an additional hydrogen on each are treated with the standard SAPT0 method in the same aDZ--aQZ bases.
The resulting energy contributions are presented in Fig.~\ref{fig:C75_frag-dimer},
showing that the intermolecular SAPT0 electrostatic energy well matches with the improved ISAPT0 variants, especially SIAO1. 
The close distance between the two capping hydrogens results in larger exchange and dispersion energies observed in the intermolecular system as expected. 
This attribution of increased first-order exchange and dispersion can be confirmed by an intermolecular F-SAPT calculation \cite{Parrish:14b}, where each SAPT term is partitioned into contributions from a pair of fragments, the capping hydrogen and the rest of the molecule.
In the F-SAPT calculation in the aTZ basis, the first-order exchange interaction energy of the added hydrogens and the partner monomer amounts to 1.71 kcal/mol, that is, 72$\%$ of the total dimer exchange energy of 2.36 kcal/mol. 
The uncapped fragments account for only 0.65 kcal/mol (28$\%$) of exchange and $-0.69$ kcal/mol (48$\%$) of dispersion energy, much closer to the intramolecular SAPT results. 
Overall, the improved ISAPT methods provide reasonable and reliable energy decomposition unlike the original ISAPT0(C) variant (which gives 1.42 kcal/mol for the electrostatic term in the C75 model). 

\subsection{Comparison of different SAO and SIAO variants, with and without orthogonalization}

The results presented so far led us to designate the ISAPT(SIAO1) variant as the most meaningful one for practical calculations. 
We also made extensive comparisons to the ISAPT(SAO1) variant as well as the original ISAPT(C) one.
We will now illustrate how the performance of the method is influenced by the level of self-consistency in the determination of fragment and link occupied orbitals (that is, the choice between SAO0/SAO1/SAO2 or SIAO0/SIAO1/SIAO2), and by the orthogonalization of the link IHOs to the fragment occupied spaces or lack thereof. 
We will make explicit comparisons using the n-heptane and 2,4-dimethylpentane models (C73 and C7B24, respectively), but the conclusions are transferable to the other studied systems as well:  analogous tables comparing different ISAPT ``minor variants'' for the pentanediol models are provided in the Supporting Information. 

\begin{figure}[!htb]
\begin{minipage}{1\textwidth}
\centering
\includegraphics[width=1\textwidth]{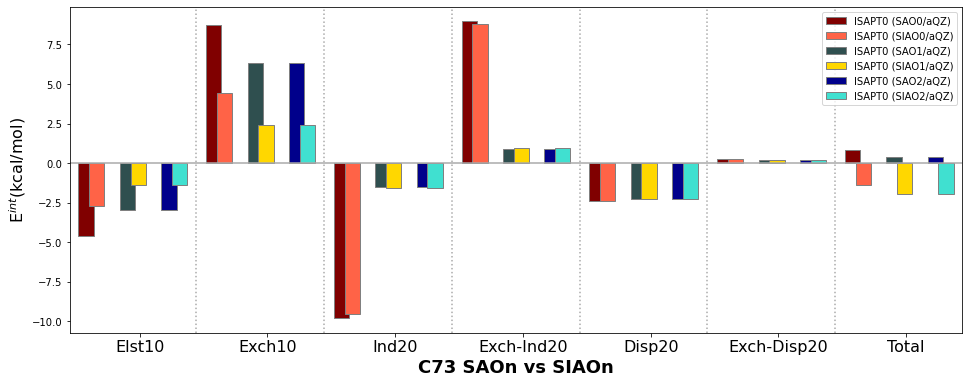}
\caption{Comparison of interaction energies in the C73 model of n-heptane, computed using the aQZ basis set with six link assignment options SAOn and SIAOn, $n=0,1,2$.}
\label{fig:C73_SAOn-SIAOn}
\end{minipage}
\end{figure}

\begin{figure}[!htb]
\begin{minipage}{1\textwidth}
\centering
\includegraphics[width=1\textwidth]{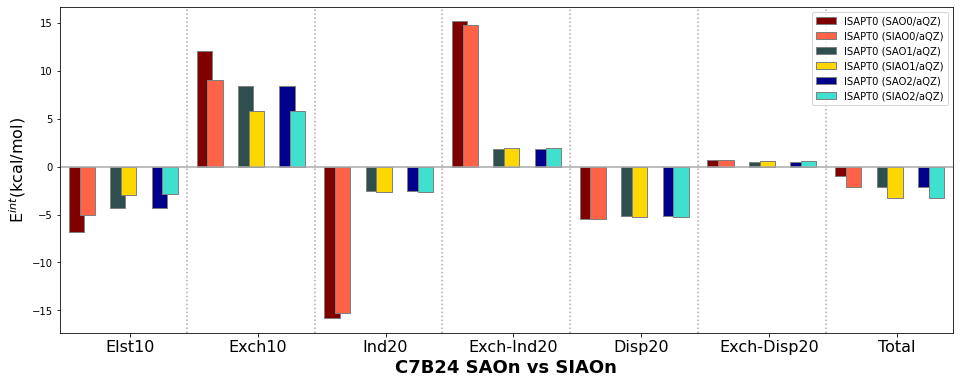}
\caption{Comparison of interaction energies in the 2,4-dimethylpentane model C7B24, computed using the aQZ basis set with six link assignment options SAOn and SIAOn, $n=0,1,2$.}
\label{fig:C7B24_SAOn-SIAOn}
\end{minipage}
\end{figure}

A comparison of the new ISAPT energy contributions for the C73 and C7B24 models is presented in Figs.~\ref{fig:C73_SAOn-SIAOn} and \ref{fig:C7B24_SAOn-SIAOn}, respectively.
In both SAO and SIAO formalisms, a dramatic change of ISAPT results, especially of induction and exchange-induction energies, is observed between the original HF orbitals (SAO0/SIAO0) and their one-iteration refinement (SAO1/SIAO1). 
However, the second iteration (SAO2/SIAO2) changes very little, confirming that the algorithm is essentially converged after one iteration. 
This is the reason why we have focused on the SAO1 and SIAO1 results so far, and why we recommend the SIAO1 variant for all practical calculations: some internal consistency between the fragment and link orbitals is clearly required (otherwise, as shown in Figs.~\ref{fig:C73_SAOn-SIAOn}-\ref{fig:C7B24_SAOn-SIAOn}, the induction effects blow up quite dramatically), but performing a single iteration of their mutual refinement is entirely sufficient.
The SAO2 and SIAO2 methods show only up to $\pm$0.003 kcal/mol energy differences compared to the corresponding SAO1 and SIAO1 methods, regardless of the basis set size.


\begin{figure}[!htb]
\begin{minipage}{\textwidth}
\centering
\includegraphics[width=\textwidth]{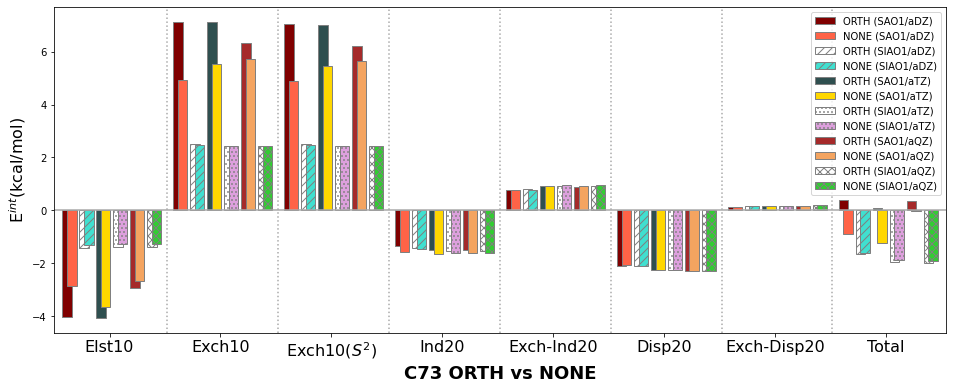}
\end{minipage}
\begin{minipage}{\textwidth}
\centering
\includegraphics[width=\textwidth]{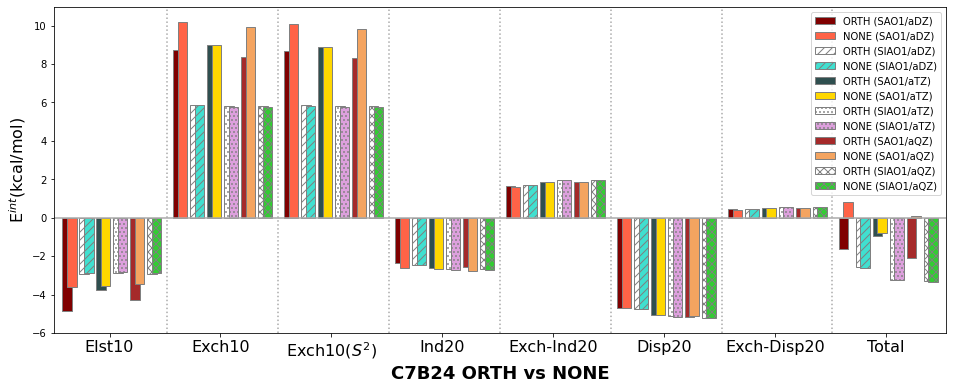}
\caption{Comparison of ISAPT energy components within n-heptane (C73) and 2,4-dimethylpentane (C7B24) molecules with (data marked ``ORTH'') and without (marked ``NONE'') link orbital orthogonalization, computed using the SAO1 and SIAO1 variants with the aXZ bases, X=D, T, Q.}
\label{fig:C73-C7B24-frag-none}
\end{minipage}
\end{figure}

Another technical detail of the new ISAPT calculations that turns out to have very minor significance is the orthogonalization of the link IHOs to the occupied space for the fragment (that is, for the SAO1 and SIAO1 approaches, the orthogonalization of $\chi'_x(\chi'_y)$ to the spaces ${\cal B}'_{\mathbf{A}}({\cal B}'_{\mathbf{B}})$, respectively).
All ISAPT(SAO) and ISAPT(SIAO) results presented so far have employed this orthogonalization, but we will now check what happens if the orthogonalization is skipped.
A comparison of the ISAPT(SAO1) and ISAPT(SIAO1) energies for n-heptane and 2,4-dimethylpentane, with (denoted ``ORTH'') and without (``NONE'') the orthogonalization of link IHOs to the fragment occupied space, is presented in Fig. \ref{fig:C73-C7B24-frag-none}.
For the SAO1 variant, the electrostatic and first-order exchange components with and without orthogonalization follow the same trends, but the actual numerical values are quite different. 
On the contrary, the differences between the corresponding first-order ISAPT(SIAO1) contributions are nearly negligible, and so are the variations in second-order ISAPT energies for both approaches.

\begin{figure}
\begin{minipage}{\textwidth}
\centering
\includegraphics[width=.6\textwidth]{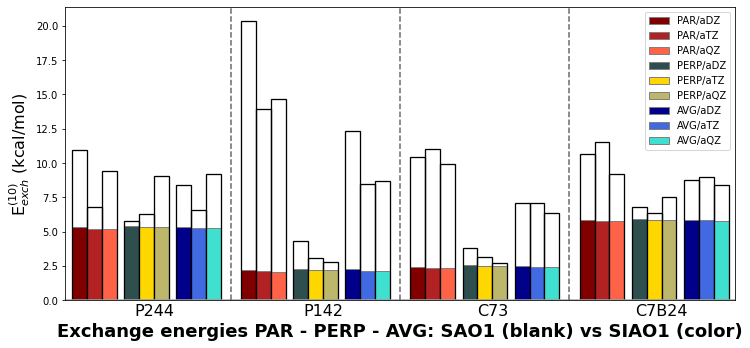}
\caption{ISAPT exchange energy $E^{(10)}_{\rm exch}$ from parallel (PAR), perpendicular (PERP), and average (AVG) spin couplings in models P244, P142, C73, and C7B24. The values are computed with SAO1 (empty bars) and SIAO1 (colored bars) in the aXZ bases, X=D,T,Q.}
\label{fig:exch10-par-perp-avg}
\end{minipage}
\end{figure}

The last technical aspect of new ISAPT calculations that we need to investigate is the choice of spin coupling between the singly occupied link IHOs in the computation of exchange energies.
At the first order, the parallel (PAR) and perpendicular (PERP) spin-coupled values of $E^{(10)}_{\rm exch}$, as well as their averages (AVG), are presented in Fig.~\ref{fig:exch10-par-perp-avg} for four systems: 2,4-pentanediol (P244), 1,4-pentanediol (P142), n-heptane (C73), and 2,4-dimethylpentane (C7B24).
Due to the larger overlap of the link orbitals, the ISAPT(SAO1) first-order exchange energy is not just more erratic with respect to basis set, but it is also much more sensitive to the spin coupling than ISAPT(SIAO1). 
As the overlap integral between the two IHOs is zero in the perpendicular mode, the parallel coupling leads to a larger exchange energy in the SAO1 variant. 
In contrast, the PAR and PERP $E^{(10)}_{\rm exch}$ values from ISAPT(SIAO1) are highly consistent for all systems studied. 
In second order (see the Supporting Information), both $E^{(20)}_{\rm exch-ind,resp}$ and $E^{(20)}_{\rm exch-disp}$  are practically independent on spin coupling: the absolute differences between the PAR and PERP values for the systems and basis sets included in Fig.~\ref{fig:exch10-par-perp-avg} do not exceed 0.042 kcal/mol (SAO1) and 2.7x10$^{-4}$ kcal/mol (SIAO1) for $E^{(20)}_{\rm exch-ind,resp}$, and 0.0032 kcal/mol (SAO1) and 7.4x10$^{-5}$ kcal/mol (SIAO1) for $E^{(20)}_{\rm exch- disp}$. For all other ISAPT results in this work, the average of the parallel and perpendicular spin-coupling values is chosen as the total exchange contribution. 

\subsection{N-arylimide molecular balances}

\begin{figure}
\begin{minipage}{.30\textwidth}
\centering
\includegraphics[angle=-15, width=.91\textwidth]{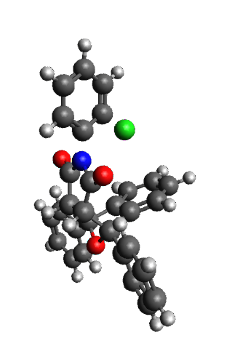}
\caption*{Cl-ben-closed}
\end{minipage}
\begin{minipage}{.30\textwidth}
\centering
\includegraphics[angle=14, width=.89\textwidth]{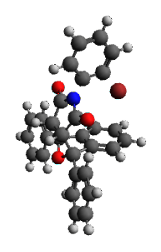}
\caption*{Br-ben-closed}
\end{minipage}
\begin{minipage}{.30\textwidth}
\centering
\includegraphics[angle=20, width=.93\textwidth]{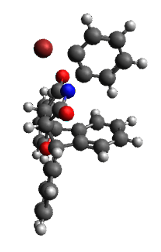}
\caption*{Br-ben-open}
\end{minipage}
\begin{minipage}{.30\textwidth}
\centering
\includegraphics[width=.65\textwidth]{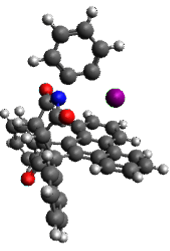}
\caption*{I-phe-closed}
\end{minipage}
\begin{minipage}{.30\textwidth}
\centering
\includegraphics[width=.65\textwidth]{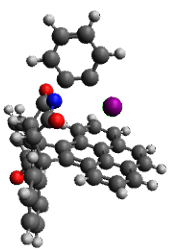}
\caption*{I-pyr-closed}
\end{minipage}
\begin{minipage}{.30\textwidth}
\centering
\includegraphics[width=.65\textwidth]{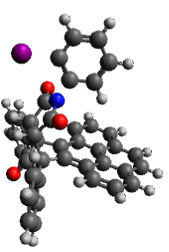}
\caption*{I-pyr-openR}
\end{minipage}
\caption{Representative open and closed structures of several molecular balances investigated in this work. The balances vary by the halogen atom (Cl, Br, or I) and its noncovalently interacting partner, the aromatic surface (ben = benzene, phe = phenanthrene, pyr = pyrene). A missing bond signifies a fragment boundary.}
\label{fig:molecular balances}
\end{figure}

Noncovalent intramolecular interactions of aromatic groups are crucial in chemical and biological processes, especially in rational drug design, new drug discovery, and the development of synthetic materials, sensors, and catalysts \cite{Sun:17,Li:20}.
A prime example of  carefully designed intramolecular interactions are the highly versatile N-arylimide molecular torsion balances, an effective platform to quantify non-covalent halogen-$\pi$ interactions in solution, via the folded (closed) and unfolded (open) conformational equilibrium. 
Recently, a number of bicyclic N-arylimide balances were synthesized and used to quantitatively probe halogen-$\pi$ interactions by Sun and coworkers \cite{Sun:17}. 
Understanding and tuning the delicate interplay of intramolecular nonbonded interactions between the open and closed conformations of molecular balances requires reliable insights from energy decomposition.
Therefore, in this study, the intramolecular interaction energy components of closed and open states of several halogen-containing N-arylimide systems (see Fig. \ref{fig:molecular balances} for representative structures, and the Supporting Information for all geometries) are analyzed using the ISAPT(SIAO1) method and compared to the original ISAPT(C) approach.
In the closed conformation, a halogen atom is positioned over an aromatic surface (benzene, phenanthrene, or pyrene). 
In the open conformation, the halogen-substituted benzene ring is rotated 180 degrees so that only a hydrogen atom points towards the aromatic surface.
It should be noted that the open and closed structures are separately optimized; thus, their geometric difference involves some rearrangements in addition to the rotation of the substituted benzene ring.
To separate the actual noncovalent intramolecular interactions from the geometry relaxation effects (the latter do not lend themselves to a SAPT-like decomposition), we also examine the ``openR'' structures, which are open-balance conformations obtained from the optimized closed structure directly by a 180-degree rotation of the halogen-substituted benzene ring (no geometry reoptimization is performed so the ``openR'' structure is not a local minimum).
An analogous ``closedR'' structure, a closed conformation obtained by a 180-degree rotation of the open one, involves very short halogen-$\pi$ distances and is strongly repulsive; therefore, we do not discuss it any further.
In all structures, the noncovalent interaction energy of the halogen atom (fragment \textbf{A}) with the bulk of the molecule including the aromatic surface (fragment \textbf{B}) is investigated, with a single phenylene ring constituting the linker fragment \textbf{C}, as depicted in Fig. \ref{fig:molecular balances}.

\sisetup{
  table-auto-round = true
}

\begin{table}
\caption{Intramolecular interaction energy components (in kcal/mol) of representative models of N-arylimide molecular balances computed at the ISAPT0/aDZ level of theory.}
\label{table:molecular balance}
\centering
\small
\nprounddigits{2}
\npdecimalsign{.}
\begin{tabular}{
    c 
    c
    S[table-format = 2.2] 
    S[table-format = 2.2] 
    S[table-format = 4.2]
    S[table-format = 3.2]
    S[table-format = 3.2]
    S[table-format = 1.2]
    S[table-format = 2.2]
  }
\hline
\multicolumn{1}{c}{\textbf{{System}}} & \multicolumn{1}{c}{\textbf{{Method}}} & \multicolumn{1}{c}{$E^{(10)}_{\rm elst}$} & \multicolumn{1}{c}{ $E^{(10)}_{\rm exch}$} & \multicolumn{1}{c}{$E^{(20)}_{\rm ind,resp}$} & \multicolumn{1}{c}{$E^{(20)}_{\rm exch-ind,resp}$} &
\multicolumn{1}{c}{$E^{(20)}_{\rm disp}$} & \multicolumn{1}{c}{$E^{(20)}_{\rm exch-disp}$} & \multicolumn{1}{c}{\textbf{{Total}}} \\ [0.5ex]
\hline\hline
\multirow{2}{*}{Cl-ben-closed} & 
{C} & -1.05766376 & 14.50441396 & -72.82229052 & 70.96901158 & -7.48981189 & 1.31967202 & 5.423331360000007 \\ 
\cline{2-9}
& {SIAO1} & -1.86711604 & 6.81755372 &  -5.90135336 & 5.3216032 & -6.82930618 & 0.79630267 & -1.662315989999999 \\
\hline
\multirow{2}{*}{Cl-ben-openR} & 
{C} & 5.42837185 & 2.64467871 & -1.91725562 & 0.92350168 & -3.08635538 & 0.27277369 & 4.265714930000001\\
\cline{2-9}
& {SIAO1} & 0.49994819 & 2.39420915 &  -1.08095034 & 0.55726081 & -3.0338976 & 0.23215413 & -0.43127565999999995\\
\hline
\multirow{2}{*}{Cl-ben-open} & 
{C} & 5.63612693 & 2.26827731 & -1.68651723 & 0.76969432 & -2.82602122 & 0.22860577 & 4.39016588 \\
\cline{2-9}
& {SIAO1} & 0.62836775 & 2.0296292 & -0.84015394 & 0.38372571 & -2.77835774 & 0.19041966 & -0.3863693600000001 \\
\hline
\multirow{2}{*}{Br-ben-closed} & 
{C} & -9.81644576 & 24.73192252 & -351.00886312 & 344.46942372 & -9.26719867 & 1.94744126 & 1.056279959999997 \\
\cline{2-9}
& {SIAO1} & -2.99073816 & 8.49351241 &  -11.10308502 & 10.34692074 & -8.33603669 & 1.09225962 & -2.4971671000000004 \\
\hline
\multirow{2}{*}{Br-ben-openR} & 
{C} & 5.53768035 & 3.02118778 & -4.35851338 & 3.30858665 & -3.41275581 & 0.33865088 & 4.434836470000001 \\
\cline{2-9}
& {SIAO1} & 0.56187103 & 2.55315214 & -1.35567346 & 0.76376358 & -3.31669894 & 0.27027983 & -0.5233058200000007 \\
\hline
\multirow{2}{*}{Br-ben-open} & 
{C} & 5.68436219 & 2.96883386 & -3.87763061 & 2.86391409 & -3.32449132 & 0.32875986 & 4.643748069999999 \\ 
\cline{2-9}
& {SIAO1} & 0.58886749 & 2.52092373 & -1.26220871 & 0.71595307 & -3.23522231 & 0.26348113 & -0.4082056000000002 \\
\hline
\multirow{2}{*}{I-phe-closed} & 
{C} & -7.53502131 & 22.9267189 & -729.84346586 & 720.46274089 & -12.42047388 & 2.27396507 & -4.135536190000038 \\
\cline{2-9}
& {SIAO1} & -2.30441041 & 8.26093582 & -9.16057683 & 8.22846951 & -11.5775715 & 1.4787948 & -5.074358609999999 \\
\hline
\multirow{2}{*}{I-phe-openR} & 
{C} & 4.86780829 & 3.95544674 & -10.82781483 & 9.49636757 & -4.42985579 & 0.51574621 & 3.577698190000001 \\
\cline{2-9}
& {SIAO1} & -0.16480613 & 3.17042644 & -2.43565082 & 1.60887828 & -4.28665523 & 0.41795144 & -1.6898560200000001 \\
\hline
\multirow{2}{*}{I-phe-open} & 
{C} & 5.4950186 & 3.50283932 & -12.4432308 & 11.24646867 & -3.91073865 & 0.43074556 & 4.3211027 \\
\cline{2-9}
& {SIAO1} & 0.18803434 & 2.67266214 & -1.81509481 & 1.14483071 & -3.77350884 & 0.3339207 & -1.24915576 \\
\hline
\multirow{2}{*}{I-pyr-closed} & 
{C} & -5.90923678 & 20.95219663 & -600.21755092 & 592.42788413 & -12.46158225 & 2.15452952 & -3.0537596699999665 \\
\cline{2-9}
& {SIAO1} & -2.25771715 & 8.02003559 & -8.7402671 & 7.81550551 & -11.68985891 & 1.44743186 & -5.404870199999999 \\
\hline
\multirow{2}{*}{I-pyr-openR} & 
{C} & 4.92871668 & 4.4540515 & -10.54748778 & 9.15235142 & -4.66149906 & 0.57232778 & 3.8984605399999994 \\
\cline{2-9}
& {SIAO1} & -0.19774462 & 3.65668875 & -2.76712814 & 1.89037462 & -4.51327719 & 0.47112742 & -1.4599591600000006 \\
\hline
\multirow{2}{*}{I-pyr-open} & 
{C} & 5.49791466 & 3.70227451 & -9.84738272 & 8.63449793 & -4.0441592 & 0.46025979 & 4.4034049699999995 \\
\cline{2-9}
& {SIAO1} & 0.14848439 & 2.9255465 & -1.96660587 & 1.28074689 & -3.90978345 & 0.3655741 & -1.15603744 \\
\hline
\end{tabular}
\end{table}

The ISAPT(C) and ISAPT(SIAO1) intramolecular interaction energy contributions for several N-arylimide molecular balances are presented in Table~\ref{table:molecular balance}.
The issues of the original C variant are quite evident. First, while the open and ``openR'' conformations display a grossly overestimated repulsive electrostatics, the ISAPT(C) $E^{(10)}_{\rm elst}$ energy for closed conformations is usually overly attractive.
While this has not happened before for the systems studied in this work, it is a manifestation of the very same issue of unphysical dipole moments at the interfragment boundary: after all, these dipole moments can be aligned unfavorably or favorably.
Second, the ISAPT(C) induction and exchange-induction terms for closed conformations are dramatically large. 
When the individual contributions are so overestimated, it is hard to expect their sufficiently complete cancellation: indeed, the sum $E^{(20)}_{\rm ind,resp}+E^{(20)}_{\rm exch-ind,resp}$ is up to 12 times larger for ISAPT(C) than for ISAPT(SIAO1).
The issues with the ISAPT(C) induction energy lead to sometimes unphysically large values of the $\delta_{\rm HF}$ term -- see the Supporting Information.
Therefore, the total ISAPT molecular balance interaction energies presented in this section {\em do not} include $\delta_{\rm HF}$.

On the other hand, a switch to the SIAO1 link assigment solves all the issues of ISAPT(C). 
The induction and exchange-induction energies for closed conformations are corrected to physically meaningful values, the first-order exchange energy for the same structures is reduced about threefold and, consequently, the total intramolecular interaction becomes attractive in all cases.  
Just like for the other systems, the dispersion energies are consistent between ISAPT(C) and ISAPT(SIAO1) ($<$1 kcal/mol differences). 
The induction and exchange-induction corrections are no longer unreasonably large and their cancellation is more complete, leaving the dispersion energy as the primary binding effect for all systems and conformations.
The ISAPT(SIAO1) electrostatic energy is weakly repulsive ($<1$ kcal/mol) for most open conformations and somewhat attractive ($-1.9$ to $-3.0$ kcal/mol) for the closed ones, but neither electrostatics nor induction match the binding strength of dispersion interactions ($-2.9$ to $-3.9$ kcal/mol for the open conformations and $-6.8$ to $-11.7$ kcal/mol for the closed ones, taking into account the small exchange-dispersion effects as well).

\begin{figure}
\begin{minipage}{\textwidth}
\centering
\includegraphics[ width=.5\textwidth]{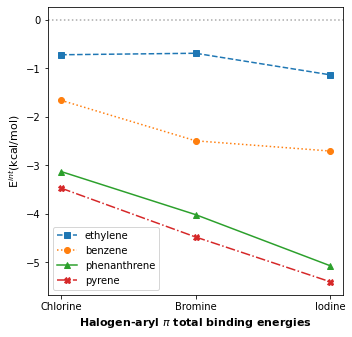}
\end{minipage}
\caption{The ISAPT0(SIAO1)/aDZ total interaction energy of closed N-arylimide molecular balances with three halogen substituents: Cl, Br, and I and four $\pi$ surfaces: ethylene, benzene, phenanthrene, and pyrene.}
\label{fig:closed total_E}
\end{figure}

Interestingly, bromine-containing systems exhibit the strongest electrostatic attraction forces in the closed configuration, ranging from $-0.71$ to $-2.99$ kcal/mol compared with $-1.14$ to $-2.31$ kcal/mol in iodine systems.
Furthermore, the benzene surface produces the strongest electrostatic energies for each halogen substituent, about 0.1 - 0.2 kcal/mol greater than in phenanthrene and pyrene systems. 
The total ISAPT(SIAO1) interaction energy in closed conformations shows that binding increases with both the size of the aromatic surface and the atomic number of the halogen (Fig. \ref{fig:closed total_E}). 
This relationship shows a good agreement with the experimental findings \cite{Sun:17}, which quantified the halogen–$\pi$ interactions to contribute about $-1.2$ kcal/mol for unimolecular systems.

For the open state of the molecular balance, the “openR” structure (where the -Ph-X fragment is rotated 180 degrees without reoptimization) has a slightly larger dihedral angle (by about 6 degrees) between the aromatic surface and the imide group connecting to the linker fragment than the “open” one. With the SIAO1 method, a slightly more repulsive first-order exchange and somewhat stronger induction and dispersion energies are observed in the “openR” systems. 
As a result, the total interaction energy of the “open” systems is up to 0.4 kcal/mol less attractive than of the “openR” systems, as shown in Table \ref{table:molecular balance}. 
However, the overall differences in ISAPT0 interaction energy components between the "open" and "openR" structures are quite minor, showing that geometric relaxation is not a crucial factor for the performance of these molecular balances; the difference in intramolecular nonbonded interactions, which can be readily studied with ISAPT(SIAO1), is a much more important quantity.
The increase in the electron density and polarizability of the halogen substituent from chlorine to iodine is correlated with their total binding energies. 
In the benzene based balances,  the overall interaction energy increases from $-0.43$ kcal/mol for chlorine to $-0.52$ kcal/mol for bromine and to $-0.98$ kcal/mol for iodine in the “openR” configurations.
The corresponding increase in the pyrene based balances is from $-0.58$ kcal/mol for Cl to $-0.75$ kcal/mol for Br and $-1.46$ kcal/mol for I.

\begin{figure}
\begin{minipage}{\textwidth}
\centering
\includegraphics[width=.5\textwidth]{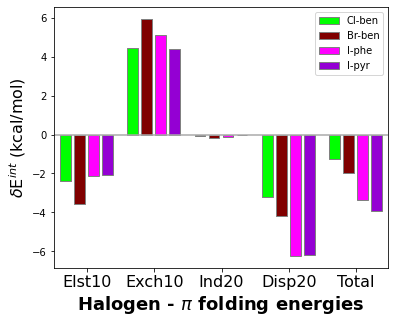}
\caption{Decomposition of the folding energy (closed$-$openR) in representative molecular balance systems computed at the ISAPT0(SIAO1)/aDZ level of theory. The effects marked ``Ind20'' and ``Disp20'' represent the {\em complete} induction and dispersion components, including the respective exchange-induction and exchange-dispersion contributions.}
\label{fig:folding-E}
\end{minipage}
\end{figure}

As shown by the ISAPT0(SIAO1) decomposition, the closed conformations of N-arylimide molecular balances are mainly bound by the dispersion forces. The predicted relative folding energy can thus be computed as $\delta E^{\rm int}_{\rm total}$(closed $-$ openR) as presented in Fig. \ref{fig:folding-E}. 
At equilibrium, due to a close halogen-$\pi$ contact, the closed structure is strongly  stabilized by dispersion forces, which grow with both the halogen atomic number (and thus its polarizability) and the size of the $\pi$ surface. 
The closed conformation is secondarily stabilized, relative to the open one, by the electrostatic contribution which is, however, more than counterbalanced by the exchange repulsion term.
The effect of induction energy on the relative conformational stability is negligible.

\section{Summary}\label{sec:summary}

In this work, we solve the issue of unphysical interaction energy contributions in the original intramolecular SAPT (ISAPT) method of Ref.~\citenum{Parrish:15}. 
This issue culminates in strongly repulsive electrostatic energies for systems where the electrostatics should clearly be favorable (such as those involving intramolecular hydrogen bonding) and while it does not show up in every possible fragmentation pattern, it plagues most of them (especially those with a small linker fragment \textbf{C}).
We identified the cause of this issue to be the artificial dipole moments at the \textbf{A}-\textbf{C} and \textbf{B}-\textbf{C} boundaries, where the linking atom is missing electrons on one of its hybrid orbitals as the entire doubly occupied intrinsic bond orbital (IBO) is assigned to \textbf{C}.
We propose to overcome this issue by partially undoing the coupling that leads from intrinsic hybrid orbitals (IHOs) to IBOs, and reassign one electron from \textbf{C} to each of \textbf{A}/\textbf{B}, placing it on a suitably constructed approximation to the missing hybrid orbital.
This reassignment leads to new, updated density matrices for fragments \textbf{A} and \textbf{B} which can be employed in standard SAPT0 formulas to compute updated electrostatic, induction, and dispersion energies (it is assumed that the link hybrid orbital does not participate in excitations).
For the corresponding exchange corrections, a choice needs to be made whether the single electrons on the \textbf{A}-\textbf{C} and \textbf{B}-\textbf{C} link orbitals undergo parallel, perpendicular, or averaged spin coupling.
This is a purely technical issue that has a minor effect on the ISAPT energy contributions.

The selection of the link IHOs involves a projection of the bond orbital onto a suitably designed fragment \textbf{A}/\textbf{B} space.
We propose two algorithms for this projection: in the SAO variant, the target space is spanned by the AO basis functions centered on \textbf{A}/\textbf{B}, and in the SIAO one, the target space is spanned by the intrinsic atomic orbitals \cite{Knizia:13} centered on \textbf{A}/\textbf{B}.
Each of these two choices leads to several minor variants depending on the level of self-consistency between the noninteracting orbital space of, say, \textbf{A} (which should not include any interaction with the \textbf{B} link IHO) and the \textbf{A} link IHO (which should be orthogonal to the other occupied orbitals of \textbf{A}).
We find out that a minimal level of this self-consistency is essential (the SAO0 and SIAO0 variants are not reliable), but one iteration towards self-consistency, as in the SAO1 and SIAO1 methods, is entirely sufficient.

We show that both ISAPT(SAO1) and ISAPT(SIAO1) resolve the issue with artificial dipole moments on the linking atoms. 
All ISAPT energy  contributions now exhibit physically meaningful values for all fragmentation schemes. 
In particular, the electrostatic energy is now strongly attractive for intramolecular hydrogen-bonded systems and weakly attractive (due to charge penetration) for nonpolar fragments.
The induction and exchange-induction energies are not nearly as large as in original ISAPT(C), which makes their sum more meaningful. 
The first-order exchange energy typically becomes less repulsive, and the dispersion and exchange-dispersion contributions are in good agreement with the original ISAPT(C) method. 
Finally, the $\delta_{\rm HF}$ term, the only component that is not amenable to the proposed link reassignment, is taken from the original ISAPT(C) theory.
In all tested molecular systems, ISAPT(SIAO1) is the best variant for calculating and interpreting intramolecular interactions. 
Unlike ISAPT(SAO1), the SIAO1 variant shows remarkable consistency between basis sets thanks to a much smaller overlap between the link IHOs when they are formed via a projection onto a fragment IAO  space. 
While both ISAPT(SAO1) and ISAPT(SIAO1) significantly reduce the magnitude of the fragment dipole moments in systems with entirely nonpolar fragments, the former variant leads to a somewhat larger reduction. 

The reliable intramolecular energy decomposition provided by the new ISAPT(SIAO1) approach enables one to shed light on the origins of various nonbonded interactions.
In this work, we illustrate the new algorithms by comparing the energy components for three pentanediol isomers at a range of fragmentation patterns, comparing the ISAPT description of the intramolecular hydrogen bond in 2,4-pentanediol to the SAPT0 description of the intermolecular hydrogen bond in the water dimer.
Next, we use ISAPT(SIAO1) to study alkanes, explaining why the branched isomer, 2,4-dimethylpentane, is more thermodynamically stable than the linear isomer, n-heptane.
Finally, we investigate a family of N-arylimide molecular torsion balances, differing by the halogen atom and the aromatic surface, to examine the origins of the folding energy, that is, the nonbonded energy difference between the closed and open conformer.
The proposed ISAPT(SIAO1) approach is expected to provide a physically reasonable energy decomposition for any closed-shell molecule that can be separated into two noncovalently interacting fragments and a linker by cutting two single bonds.
Therefore, we expect more valuable insights obtained from this method to emerge in the near future. 

\section*{Associated Content}

The Supporting Information is available free of charge at ...
\begin{itemize} 
    \item Derivation of the modified $E^{(20)}_{\rm exch-disp}$ expressions involving parallel and perpendicular spin couplings of link orbitals, and additional tables and figures (PDF)
    \item Cartesian coordinates of all systems studied in this work (TXT)
\end{itemize}

\noindent The modified Psi4 code including new ISAPT variants is available at \\ \verb+https://github.com/konpat/psi4/tree/isapt+.

\section*{Acknowledgments}

This work was supported by the U.S. National Science Foundation award CHE-1955328.

\section*{Appendix: Nonapproximated first-order exchange energy in new ISAPT}

In this Appendix, we will rederive the expression for the full, nonapproximated $E^{(10)}_{\rm exch}$ \cite{Jeziorski:76} ISAPT correction appropriate for the new SAO and SIAO variants where the singly occupied IHOs on fragments \textbf{A} and \textbf{B} do not have a definite spin.
The original formula for this correction is given in terms of spinorbitals and still holds \cite{Schaffer:12}, however, the despinning of this formula needs to proceed differently as the link spinorbitals involve both spin-up and spin-down contributions and thus couple the two spin blocks of the overlap matrix.

Following the formulation in Ref.~\citenum{Schaffer:12}, the complete nonexpanded first-order SAPT0 energy $E^{(10)}=E^{(10)}_{\rm elst}+E^{(10)}_{\rm exch}$ is expressed as
\begin{equation}
E^{(10)} = W_{AB}+\sum_{ir}B_{ir}D_{ri}+\sum_{jr}A_{jr}D_{rj}
+\frac{1}{2}\sum_{ijrs}\langle ij||rs\rangle
\left( D_{ri}D_{sj}-D_{si}D_{rj} \right)
\label{eq:e10full}
\end{equation}
In Eq.~(\ref{eq:e10full}), $i$ runs over the occupied spinorbitals of \textbf{A}, $j$ runs over occupied spinorbitals of \textbf{B}, and $r,s$ run over occupied spinorbitals of both fragments.
Furthermore, $W_{AB}$ is the constant intermolecular nuclear repulsion term, $B_{ir}=\langle \psi_i|V_B|\psi_r\rangle$ and $A_{jr}=\langle \psi_j|V_A|\psi_r\rangle$ are the matrix elements of the nuclear attraction potential of B/A, and $\langle ij||rs\rangle=\langle ij|rs\rangle-\langle ij|sr\rangle$ is the antisymmetrized two-electron integral in the physicists’ notation.
Finally, $D_{rs}$ are the elements of the inverse of the overlap matrix $S_{rs}$ of all occupied spinorbitals, ${\mathbf D}={\mathbf S}^{-1}$. 
It is the lack of the block diagonal character of ${\mathbf S}$, and thus also ${\mathbf D}$, due to spin-up and spin-down coupling that necessitates a somewhat different treatment of Eq.~(\ref{eq:e10full}) in the ISAPT/SAO and ISAPT/SIAO cases. 

Once again, the explicit form of ${\mathbf S}$ and ${\mathbf D}$ depends to some extent on the spin coupling between the electrons occupying the spinorbitals $\psi_x$ and $\psi_y$ reassigned to fragments $\mathbf{A}$ and $\mathbf{B}$ in the SAO and SIAO approaches. 
We will consider both the parallel (Eq.~(\ref{eq:psixypar})) and perpendicular (Eq.~(\ref{eq:psixyperp})) spin coupling of $\psi_x$ and $\psi_y$.
Let us assume that the occupied $\mathbf{A}$ and $\mathbf{B}$ spinorbitals $\psi_r$ are ordered as ($\psi_i(|\uparrow\rangle)$, $\psi_i(|\downarrow\rangle)$,
$\psi_x$, $\psi_j(|\uparrow\rangle)$, $\psi_j(|\downarrow\rangle)$, $\psi_y$). In this basis, the matrix $\mathbf{S}$ has the following explicit block form for each spin coupling:
\begin{equation}
\mathbf{S}^{\parallel}=\left[\begin{array}{cccccc}
\delta_{ii'} &0&0&S_{ji}&0&S_{yi}/\sqrt{2} \\
0& \delta_{ii'} &0&0&S_{ji}&S_{yi}/\sqrt{2}\\
0&0&1&S_{jx}/\sqrt{2}&S_{jx}/\sqrt{2}&S_{yx} \\
S_{ij}&0&S_{xj}/\sqrt{2}&\delta_{jj'}&0&0 \\
0&S_{ij}&S_{xj}/\sqrt{2}&0&\delta_{jj'}&0 \\
S_{iy}/\sqrt{2}&S_{iy}/\sqrt{2}&S_{xy}&0&0&1 \\
\end{array}\right]
\label{eq:smatrixpar}
\end{equation}
\begin{equation}
\mathbf{S}^{\perp}=\left[\begin{array}{cccccc}
\delta_{ii'} &0&0&S_{ji}&0&S_{yi}/\sqrt{2} \\
0& \delta_{ii'} &0&0&S_{ji}&-S_{yi}/\sqrt{2}\\
0&0&1&S_{jx}/\sqrt{2}&S_{jx}/\sqrt{2}&0 \\
S_{ij}&0&S_{xj}/\sqrt{2}&\delta_{jj'}&0&0 \\
0&S_{ij}&S_{xj}/\sqrt{2}&0&\delta_{jj'}&0 \\
S_{iy}/\sqrt{2}&-S_{iy}/\sqrt{2}&0&0&0&1 \\
\end{array}\right]
\label{eq:smatrixperp}
\end{equation}

One can see that, in either case, $\mathbf{S}$ does not have a block-diagonal character and thus needs to be inverted as a whole (therefore, a $(2N_{occ,\mathbf{A}}+2N_{occ,\mathbf{B}}+2)\times (2N_{occ,\mathbf{A}}+2N_{occ,\mathbf{B}}+2)$ matrix needs to be inverted instead of a $(N_{occ,\mathbf{A}}+N_{occ,\mathbf{B}})\times (N_{occ,\mathbf{A}}+N_{occ,\mathbf{B}})$ one in original ISAPT).
However, $\mathbf{S}$ and thus $\mathbf{D}$ remains symmetric (or, in the case of $\mathbf{S}^{\perp}$ and the $\psi_y$ row/column, antisymmetric) with respect to a simultaneous flipping of all spins. 
As a result, the corresponding (up,up) and (down,down) elements of $\mathbf{D}^{\parallel}$ or $\mathbf{D}^{\perp}$ are identical, e.g. $D^{\parallel}_{i\uparrow,j\uparrow}=D^{\parallel}_{i\downarrow,j\downarrow}=D_{ij}^{\parallel,ss}$; similarly, $D^{\parallel}_{i\uparrow,j\downarrow}=D^{\parallel}_{i\downarrow,j\uparrow}=D_{ij}^{\parallel,os}$, with the superscripts `ss' and `os' indicating same-spin and opposite-spin terms, respectively.
Now, going back to Eq.~(\ref{eq:e10full}), we note that the $B_{ir}$ and $A_{jr}$ matrices are spin-diagonal so that only the same-spin block of $\mathbf{D}$ contributes to these terms.
To despin the last term in Eq.~(\ref{eq:e10full}), we have to break up the antisymmetrized two-electron integral and note that the $\langle ij|rs\rangle$ term requires the same spins within the $(i,r)$ and $(j,s)$ pairs while the $\langle ij|sr\rangle$ term imposes the same spins within the $(i,s)$ and $(j,r)$ pairs. 
As a result, the following non-zero terms contribute to the last sum in Eq.~(\ref{eq:e10full}) (not including the $1/2$ factor): 
\begin{align}
    \sum_{ijrs}\langle ij||rs\rangle
\left( D_{ri}D_{sj}-D_{si}D_{rj} \right) =&
4\sum_{ijrs}\langle ij|rs\rangle D_{ri}^{ss}D_{sj}^{ss}
-2\sum_{ijrs}\langle ij|sr\rangle D_{ri}^{ss}D_{sj}^{ss}
\nonumber \\ &
-2\sum_{ijrs}\langle ij|sr\rangle D_{ri}^{os}D_{sj}^{os}
-2\sum_{ijrs}\langle ij|rs\rangle D_{si}^{ss}D_{rj}^{ss}
\nonumber \\ &
-2\sum_{ijrs}\langle ij|rs\rangle D_{si}^{os}D_{rj}^{os}
+4\sum_{ijrs}\langle ij|sr\rangle D_{si}^{ss}D_{rj}^{ss}
\label{eq:2elmain}
\end{align}
Special care needs to be taken when any of the indices $i,j,r,s$ falls on the reassigned link spinorbital ($x$ or $y$). Such an orbital contributes (with a factor of $1/\sqrt{2}$) to both spin contributions of an integral, and the spin-flip symmetry of $\mathbf{D}$ is in general different for the parallel and perpendicular spin coupling. 
The $\mathbf{D}^{\parallel}$ matrix is symmetric in all cases, for example, $D^{\parallel}_{x,i\uparrow}=D^{\parallel}_{x,i\downarrow}$. 
On the other hand, $\mathbf{D}^{\perp}$ is symmetric when it comes to $x$ $(D^{\perp}_{x,i}\coloneqq D^{\perp}_{x,i\uparrow}=D^{\perp}_{x,i\downarrow})$ but antisymmetric when $y$ is involved $(D^{\perp}_{y,i}\coloneqq D^{\perp}_{y,i\uparrow}=-D^{\perp}_{y,i\downarrow})$; in particular, $D^{\perp}_{xy}=0$.

The one- and two-electron integrals involving the link orbital $\psi_y$ also differ between the parallel and perpendicular spin coupling.
In the parallel case, the $\sum_{ir}B_{ir}D_{ri}$ term has the following contributions from the link orbitals, when the index $i$ falls on $x$ and/or the index $r$ falls on $x$ or $y$:
\begin{align}
\sum_{ir}B^{\parallel}_{ir}D^{\parallel}_{ri}\leadsto&
2\sum_{ii'}B_{ii'}D_{i'i}^{ss}+\sqrt{2}\sum_{i}B_{ix}D_{xi}
+2\sum_{ij}B_{ij}D_{ji}^{ss}+\sqrt{2}\sum_{i}B_{iy}D_{yi}
\nonumber \\ &
+\sqrt{2}\sum_{i}B_{xi}D_{ix}+B_{xx}D_{xx}
+\sqrt{2}\sum_{j}B_{xj}D_{jx}+B_{xy}D_{yx}
\label{eq:BDlinkpar}
\end{align}
In the perpendicular case, the same term becomes
\begin{align}
\sum_{ir}B^{\perp}_{ir}D^{\perp}_{ri}\leadsto&
2\sum_{ii'}B_{ii'}D_{i'i}^{ss}+\sqrt{2}\sum_{i}B_{ix}D_{xi}
+2\sum_{ij}B_{ij}D_{ji}^{ss}+\sqrt{2}\sum_{i}B_{iy}D_{yi}
\nonumber \\ &
+\sqrt{2}\sum_{i}B_{xi}D_{ix}+B_{xx}D_{xx}
+\sqrt{2}\sum_{j}B_{xj}D_{jx}
\label{eq:BDlinkperp}
\end{align}
with the $B_{xy}$ term canceling out in this case.
The other one-electron term is completely analogous for the parallel spin coupling, and for the perpendicular one,
once again, the term containing $A_{yx}$ cancels (note that, numerically, the parallel and perpendicular formulas are different in every term because the matrices $\mathbf{S}$ and $\mathbf{D}$ are different).
We see that every link index reduces the term prefactor by a factor of $\sqrt{2}$, in accordance with the weight of the spin-up and spin-down contributions to the overall link spinorbital. 
The same pattern is observed for the two-electron terms in Eq.~(\ref{eq:2elmain}), as will be illustrated by a complete decomposition of one of the terms involving both same-spin and opposite-spin contributions.
Omitting the summation signs for brevity, we get
\begin{small}
\begin{align}
\langle ij|sr\rangle D_{ri}D_{sj}&\leadsto 
2\langle ij|i'i''\rangle D_{i''i}^{ss}D_{i'j}^{ss}+
2\langle ij|i'i''\rangle D_{i''i}^{os}D_{i'j}^{os}+
2\langle ij|j'i'\rangle D_{i'i}^{ss}D_{j'j}^{ss}+
2\langle ij|j'i'\rangle D_{i'i}^{os}D_{j'j}^{os}
\nonumber \\ &
+\sqrt{2}\langle ij|xi'\rangle D_{i'i}^{ss}D_{xj}+
\sqrt{2}\langle ij|xi'\rangle D_{i'i}^{os}D_{xj}+
\sqrt{2}\langle ij|yi'\rangle D_{i'i}^{ss}D_{yj}\pm
\sqrt{2}\langle ij|yi'\rangle D_{i'i}^{os}D_{yj}
\nonumber \\ &
+2\langle ij|i'j'\rangle D_{j'i}^{ss}D_{i'j}^{ss}+
2\langle ij|i'j'\rangle D_{j'i}^{os}D_{i'j}^{os}+
2\langle ij|j'j''\rangle D_{j''i}^{ss}D_{j'j}^{ss}+
2\langle ij|j'j''\rangle D_{j''i}^{os}D_{j'j}^{os}
\nonumber \\ &
+\sqrt{2}\langle ij|xj'\rangle D_{j'i}^{ss}D_{xj}+
\sqrt{2}\langle ij|xj'\rangle D_{j'i}^{os}D_{xj}+
\sqrt{2}\langle ij|yj'\rangle D_{j'i}^{ss}D_{yj}\pm
\sqrt{2}\langle ij|yj'\rangle D_{j'i}^{os}D_{yj}
\nonumber \\ &
+\sqrt{2}\langle ij|i'x\rangle D_{xi}D_{i'j}^{ss}+
\sqrt{2}\langle ij|i'x\rangle D_{xi}D_{i'j}^{os}+
\sqrt{2}\langle ij|j'x\rangle D_{xi}D_{j'j}^{ss}+
\sqrt{2}\langle ij|j'x\rangle D_{xi}D_{j'j}^{os}
\nonumber \\ &
+2\langle ij|xx\rangle D_{xi}D_{xj}
\underline{+2\langle ij|yx\rangle D_{xi}D_{yj}}
\nonumber \\ &
+\sqrt{2}\langle ij|i'y\rangle D_{yi}D_{i'j}^{ss}\pm
\sqrt{2}\langle ij|i'y\rangle D_{yi}D_{i'j}^{os}+
\sqrt{2}\langle ij|j'y\rangle D_{yi}D_{j'j}^{ss}\pm
\sqrt{2}\langle ij|j'y\rangle D_{yi}D_{j'j}^{os}
\nonumber \\ &
\underline{+2\langle ij|xy\rangle D_{yi}D_{xj}}
+2\langle ij|yy\rangle D_{yi}D_{yj}
\nonumber \\ &
+\sqrt{2}\langle xj|i'i\rangle D_{ix}D_{i'j}^{ss}+
\sqrt{2}\langle xj|i'i\rangle D_{ix}D_{i'j}^{os}+
\sqrt{2}\langle xj|j'i\rangle D_{ix}D_{j'j}^{ss}+
\sqrt{2}\langle xj|j'i\rangle D_{ix}D_{j'j}^{os}
\nonumber \\ &
+2\langle xj|xi\rangle D_{ix}D_{xj}
\underline{+2\langle xj|yi\rangle D_{ix}D_{yj}}
\nonumber \\ &
+\sqrt{2}\langle xj|ij'\rangle D_{j'x}D_{ij}^{ss}+
\sqrt{2}\langle xj|ij'\rangle D_{j'x}D_{ij}^{os}+
\sqrt{2}\langle xj|j'j''\rangle D_{j''x}D_{j'j}^{ss}+
\sqrt{2}\langle xj|j'j''\rangle D_{j''x}D_{j'j}^{os}
\nonumber \\ &
+2\langle xj|xj'\rangle D_{j'x}D_{xj}
\underline{+2\langle xj|yj'\rangle D_{j'x}D_{yj}}
\nonumber \\ &
+\langle xj|ix\rangle D_{xx}D_{ij}^{ss}+
\langle xj|ix\rangle D_{xx}D_{ij}^{os}+
\langle xj|j'x\rangle D_{xx}D_{j'j}^{ss}+
\langle xj|j'x\rangle D_{xx}D_{j'j}^{os}
\nonumber \\ &
+\sqrt{2}\langle xj|xx\rangle D_{xx}D_{xj}
\underline{+\sqrt{2}\langle xj|yx\rangle D_{xx}D_{yj}}
\nonumber \\ &
\underline{+\langle xj|iy\rangle D_{yx}D_{ij}^{ss}+
\langle xj|iy\rangle D_{yx}D_{ij}^{os}+
\langle xj|j'y\rangle D_{yx}D_{j'j}^{ss}+
\langle xj|j'y\rangle D_{yx}D_{j'j}^{os}}
\nonumber \\ &
\underline{+\sqrt{2}\langle xj|xy\rangle D_{yx}D_{xj}+
\sqrt{2}\langle xj|yy\rangle D_{yx}D_{yj}}
\nonumber \\ &
+\sqrt{2}\langle iy|i'i''\rangle D_{i''i}^{ss}D_{i'y}\pm
\sqrt{2}\langle iy|i'i''\rangle D_{i''i}^{os}D_{i'y}+
\sqrt{2}\langle iy|ji'\rangle D_{i'i}^{ss}D_{jy}\pm
\sqrt{2}\langle iy|ji'\rangle D_{i'i}^{os}D_{jy}
\nonumber \\ &
\underline{+\langle iy|xi'\rangle D_{i'i}^{ss}D_{xy}+
\langle iy|xi'\rangle D_{i'i}^{os}D_{xy}}
+\langle iy|yi'\rangle D_{i'i}^{ss}D_{yy}\pm
\langle iy|yi'\rangle D_{i'i}^{os}D_{yy}
\nonumber \\ &
+\sqrt{2}\langle iy|i'j\rangle D_{ji}^{ss}D_{i'y}\pm
\sqrt{2}\langle iy|i'j\rangle D_{ji}^{os}D_{i'y}+
\sqrt{2}\langle iy|j'j\rangle D_{ji}^{ss}D_{j'y}\pm
\sqrt{2}\langle iy|j'j\rangle D_{ji}^{os}D_{j'y}
\nonumber \\ &
\underline{+\langle iy|xj\rangle D_{ji}^{ss}D_{xy}+
\langle iy|xj\rangle D_{ji}^{os}D_{xy}}
+\langle iy|yj\rangle D_{ji}^{ss}D_{yy}\pm
\langle iy|yj\rangle D_{ji}^{os}D_{yy}
\nonumber \\ &
\underline{+2\langle iy|i'x\rangle D_{xi}D_{i'y}+
2\langle iy|jx\rangle D_{xi}D_{jy}+
\sqrt{2}\langle iy|xx\rangle D_{xi}D_{xy}+
\sqrt{2}\langle iy|yx\rangle D_{xi}D_{yy}}
\nonumber \\ &
+2\langle iy|i'y\rangle D_{yi}D_{i'y}+
2\langle iy|jy\rangle D_{yi}D_{jy}
\underline{+\sqrt{2}\langle iy|xy\rangle D_{yi}D_{xy}}
+\sqrt{2}\langle iy|yy\rangle D_{yi}D_{yy}
\nonumber \\ &
\underline{+2\langle xy|i'i\rangle D_{ix}D_{i'y}+
2\langle xy|ji\rangle D_{ix}D_{jy}
+\sqrt{2}\langle xy|xi\rangle D_{ix}D_{xy}
+\sqrt{2}\langle xy|yi\rangle D_{ix}D_{yy}}
\nonumber \\ &
\underline{+2\langle xy|ij\rangle D_{jx}D_{iy}+
2\langle xy|j'j\rangle D_{jx}D_{j'y}
+\sqrt{2}\langle xy|xj\rangle D_{jx}D_{xy}+
\sqrt{2}\langle xy|yj\rangle D_{jx}D_{yy}}
\nonumber \\ &
\underline{+\sqrt{2}\langle xy|ix\rangle D_{xx}D_{iy}+
\sqrt{2}\langle xy|jx\rangle D_{xx}D_{jy}+
\langle xy|xx\rangle D_{xx}D_{xy}+
\langle xy|yx\rangle D_{xx}D_{yy}}
\nonumber \\ &
\underline{+\sqrt{2}\langle xy|iy\rangle D_{yx}D_{iy}+
\sqrt{2}\langle xy|jy\rangle D_{yx}D_{jy}+
\langle xy|xy\rangle D_{yx}D_{xy}+
\langle xy|yy\rangle D_{yx}D_{yy}}
\label{eq:2ellink}    
\end{align}
\end{small}
We see that each occurrence of a link-orbital index $x$ or $y$ simply lowers the prefactor of a term by a factor of $\sqrt{2}$, from a total of 4 (twice the same-spin part plus twice the opposite-spin part) when no link indices are present to 1 when all four indices pertain to link orbitals.
The same behavior occurs for all other two-electron terms. 
The variant of Eq.~(\ref{eq:2ellink}) for the perpendicular spin coupling omits all terms containing $D^{\perp}_{xy}=D^{\perp}_{yx}=0$, and also the two-electron integral is zero if $x$ and $y$ are the pair of indices belonging to the same electron, e.g., $\langle xj|yi\rangle^{\perp}=0$. As a result, all terms underlined in Eq.~(\ref{eq:2ellink}) are present only in the parallel spin coupling and they are omitted in the perpendicular spin coupling variant. 
Moreover, the terms involving both $y$ and the opposite-spin $D$ matrix elements change sign; thus the $\pm$ signs in Eq.~(\ref{eq:2ellink}) which signify a plus sign for the parallel spin coupling and a minus sign for the perpendicular one.
An analogous formula for a Coulomb-type integral term, once again in the parallel and perpendicular version, reads
\begin{align}
\langle ij|rs\rangle D_{ri}D_{sj}&\leadsto 
4\langle ij|i''i'\rangle D_{i''i}^{ss}D_{i'j}^{ss}+
4\langle ij|i'j'\rangle D_{i'i}^{ss}D_{j'j}^{ss}+
2\sqrt{2}\langle ij|i'x\rangle D_{i'i}^{ss}D_{xj}+
2\sqrt{2}\langle ij|i'y\rangle D_{i'i}^{ss}D_{yj}
\nonumber \\ &
+4\langle ij|j'i'\rangle D_{j'i}^{ss}D_{i'j}^{ss}+
4\langle ij|j''j'\rangle D_{j''i}^{ss}D_{j'j}^{ss}+
2\sqrt{2}\langle ij|j'x\rangle D_{j'i}^{ss}D_{xj}+
2\sqrt{2}\langle ij|j'y\rangle D_{j'i}^{ss}D_{yj}
\nonumber \\ &
+2\sqrt{2}\langle ij|xi'\rangle D_{xi}D_{i'j}^{ss}+
2\sqrt{2}\langle ij|xj'\rangle D_{xi}D_{j'j}^{ss}+
2\langle ij|xx\rangle D_{xi}D_{xj}+
2\langle ij|xy\rangle D_{xi}D_{yj}
\nonumber \\ &
+2\sqrt{2}\langle ij|yi'\rangle D_{yi}D_{i'j}^{ss}+
2\sqrt{2}\langle ij|yj'\rangle D_{yi}D_{j'j}^{ss}+
2\langle ij|yx\rangle D_{yi}D_{xj}+
2\langle ij|yy\rangle D_{yi}D_{yj}
\nonumber \\ &
+2\sqrt{2}\langle xj|ii'\rangle D_{ix}D_{i'j}^{ss}+
2\sqrt{2}\langle xj|ij'\rangle D_{ix}D_{j'j}^{ss}+
2\langle xj|ix\rangle D_{ix}D_{xj}+
2\langle xj|iy\rangle D_{ix}D_{yj}
\nonumber \\ &
+2\sqrt{2}\langle xj|j'i\rangle D_{j'x}D_{ij}^{ss}+
2\sqrt{2}\langle xj|j''j'\rangle D_{j''x}D_{j'j}^{ss}+
2\langle xj|j'x\rangle D_{j'x}D_{xj}+
2\langle xj|j'y\rangle D_{j'x}D_{yj}
\nonumber \\ &
+2\langle xj|xi\rangle D_{xx}D_{ij}^{ss}+
2\langle xj|xj'\rangle D_{xx}D_{j'j}^{ss}+
\sqrt{2}\langle xj|xx\rangle D_{xx}D_{xj}+
\sqrt{2}\langle xj|xy\rangle D_{xx}D_{yj}
\nonumber \\ &
\underline{+2\langle xj|yi\rangle D_{yx}D_{ij}^{ss}+
2\langle xj|yj'\rangle D_{yx}D_{j'j}^{ss}+
\sqrt{2}\langle xj|yx\rangle D_{yx}D_{xj}+
\sqrt{2}\langle xj|yy\rangle D_{yx}D_{yj}}
\nonumber \\ &
+2\sqrt{2}\langle iy|i''i'\rangle D_{i''i}^{ss}D_{i'y}+
2\sqrt{2}\langle iy|i'j\rangle D_{i'i}^{ss}D_{jy}
\underline{+2\langle iy|i'x\rangle D_{i'i}^{ss}D_{xy}}
+2\langle iy|i'y\rangle D_{i'i}^{ss}D_{yy}
\nonumber \\ &
+2\sqrt{2}\langle iy|ji'\rangle D_{ji}^{ss}D_{i'y}+
2\sqrt{2}\langle iy|jj'\rangle D_{ji}^{ss}D_{j'y}
\underline{+2\langle iy|jx\rangle D_{ji}^{ss}D_{xy}}
+2\langle iy|jy\rangle D_{ji}^{ss}D_{yy}
\nonumber \\ &
+2\langle iy|xi'\rangle D_{xi}D_{i'y}+
2\langle iy|xj\rangle D_{xi}D_{jy}
\underline{+\sqrt{2}\langle iy|xx\rangle D_{xi}D_{xy}}
+\sqrt{2}\langle iy|xy\rangle D_{xi}D_{yy}
\nonumber \\ &
+2\langle iy|yi'\rangle D_{yi}D_{i'y}+
2\langle iy|yj\rangle D_{yi}D_{jy}
\underline{+\sqrt{2}\langle iy|yx\rangle D_{yi}D_{xy}}
+\sqrt{2}\langle iy|yy\rangle D_{yi}D_{yy}
\nonumber \\ &
+2\langle xy|ii'\rangle D_{ix}D_{i'y}+
2\langle xy|ij\rangle D_{ix}D_{jy}
\underline{+\sqrt{2}\langle xy|ix\rangle D_{ix}D_{xy}}
+\sqrt{2}\langle xy|iy\rangle D_{ix}D_{yy}
\nonumber \\ &
+2\langle xy|ji\rangle D_{jx}D_{iy}+
2\langle xy|jj'\rangle D_{jx}D_{j'y}
\underline{+\sqrt{2}\langle xy|jx\rangle D_{jx}D_{xy}}
+\sqrt{2}\langle xy|jy\rangle D_{jx}D_{yy}
\nonumber \\ &
+\sqrt{2}\langle xy|xi\rangle D_{xx}D_{iy}+
\sqrt{2}\langle xy|xj\rangle D_{xx}D_{jy}
\underline{+\langle xy|xx\rangle D_{xx}D_{xy}}
+\langle xy|xy\rangle D_{xx}D_{yy}
\nonumber \\ &
\underline{+\sqrt{2}\langle xy|yi\rangle D_{yx}D_{iy}+
\sqrt{2}\langle xy|yj\rangle D_{yx}D_{jy}+
\langle xy|yx\rangle D_{yx}D_{xy}+
\langle xy|yy\rangle D_{yx}D_{yy}}
\label{eq:2ellinkJ}    
\end{align}
with the underlined terms omitted for the perpendicular spin coupling.

The molecular-orbital formulas for the complete, nonapproximated $E^{(10)}$ correction still need to be recast into an atomic-orbital (AO) form so that the operations involving two-electron integrals can be efficiently implemented using generalized Coulomb and exchange matrices \cite{Parrish:17}.
To this end, we will use capital letters $K,L,M,N$ to denote the AO basis functions; note that the set of AOs spans the entire molecule and is used to expand occupied orbitals on all fragments. 
The coefficient $C_{rK}$ represents the weight of basis function $K$ in the molecular orbital $\psi_r$, that is, $\psi_r=\sum_K C_{rK} \phi_K^{AO}$.
Further, we define the back-transformed inverse-overlap matrix blocks $\mathbf{D}^{vw,spin}$, where $v,w$ are orbital type indices ($i$, $j$, $r$, or $s$) and the spin case $spin$ is either `ss' or `os', as in the following example:
\begin{align}
(\mathbf{D}^{ri,ss})_{KL}="\sum_{ri} C_{rK}D_{ri}^{ss}C_{iL}"
=&\sum_{i'i} C_{i'K}D_{i'i}^{ss}C_{iL}
+\frac{1}{\sqrt{2}}\sum_{i} C_{xK}D_{xi}C_{iL}
\nonumber \\ &
+\sum_{ji} C_{jK}D_{ji}^{ss}C_{iL}
+\frac{1}{\sqrt{2}}\sum_{i} C_{yK}D_{yi}C_{iL}
\nonumber \\ &
+\frac{1}{\sqrt{2}}\sum_{i} C_{iK}D_{ix}C_{xL}
+\frac{1}{2}C_{xK}D_{xx}C_{xL}
\nonumber \\ &
+\frac{1}{\sqrt{2}}\sum_{j} C_{jK}D_{jx}C_{xL}
+\frac{1}{2}C_{yK}D_{yx}C_{xL}
\label{eq:DAOdef}
\end{align}
A second example shows the differences between the parallel spin coupling (upper signs) and perpendicular spin coupling (lower signs) in the opposite-spin case (the same-spin formulas are identical for both couplings):
\begin{align}
(\mathbf{D}^{sj,os})_{KL}="\sum_{sj} C_{sK}D_{sj}^{os}C_{jL}"
=&\sum_{ij} C_{iK}D_{ij}^{os}C_{jL}
+\frac{1}{\sqrt{2}}\sum_{i} C_{xK}D_{xj}C_{jL}
\nonumber \\ &
+\sum_{jj'} C_{jK}D_{jj'}^{os}C_{j'L}
\pm\frac{1}{\sqrt{2}}\sum_{i} C_{yK}D_{yj}C_{jL}
\nonumber \\ &
\pm\frac{1}{\sqrt{2}}\sum_{i} C_{iK}D_{iy}C_{yL}
+\frac{1}{2}C_{xK}D_{xy}C_{yL}
\nonumber \\ &
\pm\frac{1}{\sqrt{2}}\sum_{j} C_{jK}D_{jy}C_{yL}
\pm\frac{1}{2}C_{yK}D_{yy}C_{yL}
\label{eq:DAOdef2}
\end{align}
(additionally, as stated before, for the perpendicular spin coupling $D_{xy}=0$).
The intermediate expression in quotes would be appropriate in a closed-shell case where each occupied orbital of \textbf{A} and \textbf{B} holds two electrons, and it has been used before \cite{Smith:20,Waldrop:21}.
However, in our case we need to break down summations over individual-fragment indices into doubly occupied orbitals and the link orbital $(i\leadsto (i,x),\; j\leadsto (j,y))$,
and the summations over both-fragment indices into all four above parts $(r,s\leadsto (i,x,j,y))$. 
To accommodate the correct numerical prefactors for terms involving link orbitals in Eqs.~(\ref{eq:BDlinkpar})--(\ref{eq:2ellink}), each term in 
Eqs.~(\ref{eq:DAOdef})--(\ref{eq:DAOdef2}) is reduced by a factor of $\sqrt{2}$ for every link orbital $x,y$ present.
One should note that $v,w$ in $\mathbf{D}^{vw,spin}$ are not matrix indices but only denote the specific molecular orbital subblock over which the summation in Eq.~(\ref{eq:DAOdef}) extends. 
Regardless of the $v,w$ type, all these matrices are of the same size $N_{AO}\times N_{AO}$.
Finally, as implied in Eqs.~(\ref{eq:DAOdef})--(\ref{eq:DAOdef2}), the ss/os designation only applies if neither of the indices represents a link orbital.

The definition of $\mathbf{D}^{vw,spin}$ in Eqs.~(\ref{eq:DAOdef})--(\ref{eq:DAOdef2}) allows folding all contributions to any single term in Eq.~(\ref{eq:2elmain}), for example, the 64 terms in Eq.~(\ref{eq:2ellink}), into one tensor contraction.
The same is achieved for the 8 contributions in Eq.~(\ref{eq:BDlinkpar}) and the analogous 8 contributions involving the $A_{jr}$ matrix. 
This leads to the following AO formula for the complete first-order ISAPT interaction energy with the reassignment of link orbitals:
\begin{align}
E^{(10)} = &\; W_{AB}+2\mathbf{B}\cdot\mathbf{D}^{ri,ss}
+2\mathbf{A}\cdot\mathbf{D}^{rj,ss}
\nonumber \\ &
+2\mathbf{D}^{ri,ss}\cdot \mathbf{J}[\mathbf{D}^{sj,ss}]
-\mathbf{D}^{ri,ss}\cdot\mathbf{K}[\mathbf{D}^{sj,ss}]^{T}
-\mathbf{D}^{ri,os}\cdot\mathbf{K}[\mathbf{D}^{sj,os}]^{T}
\nonumber \\ &
-\mathbf{D}^{si,ss}\cdot\mathbf{K}[\mathbf{D}^{rj,ss}]^{T}
-\mathbf{D}^{si,os}\cdot\mathbf{K}[\mathbf{D}^{rj,os}]^{T}
+2\mathbf{D}^{si,ss}\cdot\mathbf{J}[\mathbf{D}^{rj,ss}]
\nonumber \\ = & \;
W_{AB}+2\mathbf{B}\cdot\mathbf{D}^{ri,ss}
+2\mathbf{A}\cdot\mathbf{D}^{rj,ss}
\nonumber \\ &
+4\mathbf{D}^{ri,ss}\cdot \mathbf{J}[\mathbf{D}^{sj,ss}]
-2\mathbf{D}^{ri,ss}\cdot\mathbf{K}[\mathbf{D}^{sj,ss}]^{T}
-2\mathbf{D}^{ri,os}\cdot\mathbf{K}[\mathbf{D}^{sj,os}]^{T}
\label{eq:e10fullao}
\end{align}
where the back-transformed inverse-overlap matrix blocks $\mathbf{D}^{vw,spin}$, augmented by link orbital contributions, are defined in Eqs.~(\ref{eq:DAOdef})--(\ref{eq:DAOdef2}), and 
the generalized Coulomb and exchange matrices 
$\mathbf{J}[\mathbf{X}]$ and $\mathbf{K}[\mathbf{X}]$ have been defined in
Eq.~(\ref{eq:JKmat}).
The final transformation in Eq.~(\ref{eq:e10fullao}) makes use of the fact that the indices $r$ and $s$ are equivalent (have the same summation range), so after summing over the MO indices, for example,
$\mathbf{D}^{ri,ss}\equiv\mathbf{D}^{si,ss}$.
Equation (\ref{eq:e10fullao}) has the same form for the parallel and perpendicular spin coupling, but the $\mathbf{D}^{vw,spin}$ matrix blocks, as explained above, are defined and computed differently.

\bibliography{master}

\clearpage

\begin{figure}
\begin{minipage}{\textwidth}
\centering
\includegraphics[ width=\textwidth]{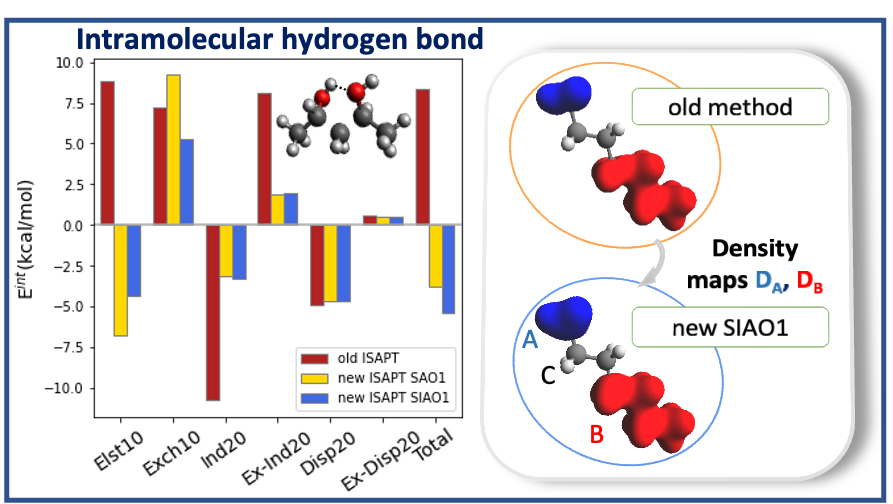}
\caption{TOC graphic}
\end{minipage}
\end{figure}

\end{document}